\documentclass[11pt]{article}
\mathchardef\mhyphen="2D 

\usepackage{lipsum} 

\usepackage{bm}
\pdfoutput=1
\usepackage{amsmath,amsfonts,amsthm}
\usepackage{esdiff}  
\usepackage{booktabs}  
\usepackage{url}  
\usepackage{cleveref}  
	\crefname{equation}{equation}{equations}
	\crefname{figure}{figure}{figures}	
	\crefname{table}{table}{tables}
\usepackage[skip=1.5pt,font=small]{caption}

\usepackage{caption}
\usepackage{bbm}

\usepackage[usenames,dvipsnames,svgnames,table]{xcolor}

\usepackage{graphicx}
\usepackage{hyperref}  

\usepackage[sc]{mathpazo} 
\usepackage[T1]{fontenc} 
\usepackage{microtype} 

\usepackage{multicol} 
\usepackage[margin={1cm,1.5cm}]{geometry}
\usepackage{booktabs} 
\usepackage{float} 
\usepackage{hyperref} 

\usepackage{lettrine} 
\usepackage{paralist} 

\usepackage{abstract} 

\usepackage{titlesec} 
\renewcommand\thesection{\Roman{section}} 
\renewcommand\thesubsection{\Alph{subsection}} 
\titleformat{\section}[block]{\large\scshape\centering\bfseries}{\thesection.}{1em}{} 

\titleformat{\subsection}[block]{\scshape\centering}{\thesubsection.}{1em}{} 

\usepackage[export]{adjustbox}
\usepackage{fancyhdr} 
\DeclareCaptionFormat{myformat}{#1#2#3\hrulefill}
\usepackage{authblk}
\makeatletter
\renewcommand\AB@affilsepx{, \protect\Affilfont}
\makeatother
\title{\vspace{-15mm}\fontsize{16pt}{16pt}\selectfont\textbf{Model of the Songbird Nucleus HVC as a Network of Central Pattern Generators}} %
\author[1]{Eve Armstrong\thanks{earmstrong@ucsd.edu}}
\author[2,3]{Henry D. I. Abarbanel\thanks{habarbanel@ucsd.edu}}
\affil[1]{BioCircuits Institute} 
\affil[2]{Department of Physics}
\affil[3]{Marine Physical Laboratory (Scripps Institution of Oceanography), University of California, San Diego, La Jolla, CA 92093-0374}
\date{(Dated: \today)\vspace{-2ex}}
\renewcommand\Affilfont{\itshape\small}

\begin{document}
\maketitle 
\begin{abstract}
We propose a functional architecture of the adult songbird nucleus HVC in which the core element is a "functional syllable unit" (FSU).  In this model, HVC is organized into FSUs, each of which provides the basis for the production of one syllable in vocalization.  Within each FSU, the inhibitory neuron population takes one of two operational states: (A) simultaneous firing wherein all inhibitory neurons fire simultaneously, and (B) competitive firing of the inhibitory neurons.  Switching between these basic modes of activity is accomplished via changes in the synaptic strengths among the inhibitory neurons.  The inhibitory neurons connect to excitatory projection neurons such that during state (A) the activity of projection neurons is suppressed, while during state (B) patterns of sequential firing of projection neurons can occur.  The latter state is stabilized by feedback from the projection to the inhibitory neurons.  Song composition for specific species is distinguished by the manner in which different FSUs are functionally connected to each other.  

Ours is a computational model built with biophysically based neurons.  We illustrate that many observations of HVC activity are explained by the dynamics of the proposed population of FSUs, and we identify aspects of the model that are currently testable experimentally.  In addition, and standing apart from the core features of an FSU, we propose that the transition between modes may be governed by the biophysical mechanism of neuromodulation. 
\end{abstract}
\newpage
\section{\\INTRODUCTION}
\begin{multicols}{2}
In the song system of the avian brain, nucleus HVC plays a central role at the junction of the auditory and song production pathways.  The sparse firing of excitatory projection neurons within HVC into the song production pathway via nucleus RA ($HVC_{RA}$ cells) has been the subject of both experiments and model development.  A critical role of the inhibitory interneurons ($HVC_I$ cells) was suggested by Gibb, Gentner, \& Abarbanel (2009a) (hereafter GGA1) and supported by experiments by Kosche et al. (2015) (hereafter KVL15).  In light of those experiments and earlier observations, we suggest an architecture for HVC that is consistent with a critical role for both structured inhibition and excitation.  This paper lays out and explores the consequences of this architecture, which is built on functional units in HVC that underly the formation of syllables in vocalization.

We model the nucleus HVC as a pattern-generating network capable of qualitatively reproducing many observations at the whole-cell, population, and behavioral levels.  In doing so, we offer a biophysical explanation for the sparse firings of HVC projection neurons established by Hahnloser et al. (2002) and the experiments that subsequently uncovered details of the underlying neuronal processes and neural circuit relationships, particularly the work of KVL15.  

We depart from previous modeling efforts by shifting focus from the sparse firings alone to a broader picture including other salient features of HVC that complementary lines of research have illuminated.  In particular, we are interested in a cell's effect upon a circuit at times when it is active and equally at times when it is \textit{not} active.  

Our model consists of two key components: i) a sub-circuit that can assume one of two modes of behavior, depending on inhibitory synaptic coupling strengths, and ii) a mechanism capable of effecting a rapid transition between these modes.  We find that these aims can be accomplished by invoking Lotka-Volterra-like dynamics subject to possible neuromodulatory mechanisms that have been proposed to explain neuronal activity in a variety of species, but which have been studied minimally within the context of the avian song generation system.  

Hahnloser et al. (2002) established the pattern-generating capability of HVC by demonstrating that some $HVC_{RA}$ neurons reliably participate in one sparse pattern during song.  Lesioning studies had previously identified HVC as residing high on the control pathway for song production, thereby poised to pass its instructions downstream where the song is ultimately generated by the syrinx and lungs.  The cellular dynamics in HVC have been significantly illuminated by KVL15, which focused on the interplay between $HVC_{RA}$ and $HVC_I$ neuronal activity.  KVL15 demonstrated via a series of GABA antagonist (gabazine) induced responses that the role of inhibition is central in modulating activity of the $HVC_{RA}$ cells.  They established the importance of both a structured inhibition and structured excitation for song generation, where all activity is set within an ambient background of excitation.  In addition, KVL15 reported high \textit{in vitro} rates of reciprocal connectivity between cell pairs and disynaptic connectivity between $HVC_{RA}$ cells, which Mooney \& Prather (2005) had noted are reminiscent of such rates in pattern-generating networks throughout the central nervous systems in other species.  Pattern-generating activity has also been induced via electrical stimulations in slice preparations of HVC (Solis \& Perkel 2005).

Previous HVC models have focused on producing the observed series of $HVC_{RA}$ neuron activations, by invoking a feedforward "synfire" chain of excitation (Li \& Greenside 2006; Long, Jin, \& Fee 2010; GGA1).  GGA1 suggested in addition that inhibition plays a mediating role upon such a chain.  One of the artificial constructs of GGA1 was the introduction of a neuronal oscillator loop that could transition between an "on" and "off" state.  This functional loop was arrayed in a chain stimulated in an unspecified manner to excite a signal moving down the chain, as interneuron activity confined the excitation to a short temporal window.  In this way, GGA1 suggested that inhibition is integral to the series propagation, yet their proposed mechanism was carefully engineered without biophysical motivations.  Cannon et al. (2015) also proposed a chain modulated by inhibition, without offering a biophysical motivation for the form of that modulation.  Moreover, while chain models can capture the propagation of series activations, the chain model is troublesome in that, by its very definition, it does not represent the picture of a strongly interconnected web - the picture that emerges from KVL15, where both structured excitation and structured inhibition play integral roles. 

An alternative to the chain model has been offered in terms of a competition among neurons.  Two basic types of competition have been applied to the modeling of neuronal networks: winner-take-all (e.g. Verduzco-Flores et al. 2012) and winnerless (e.g. Yildiz \& Kiebel 2011).  While the latter authors proposed no particular connectivity for effecting the functionality, we found their invocation of winnerless competition (WLC) appealing for the relative simplicity with which it can, in principle, generate activity reminiscent of HVC activity.  More importantly, the WLC formalism describes an identified biophysical process: the observed phenomenon of mutual inhibition.  Indeed, the framers of WLC themselves have suggested WLC as a likely underlier of series activity in HVC (Afraimovich et al. 2004; Rabinovich et al. 2006).  We have taken this idea as a critical component of the model that we propose in this paper.

We propose a model of HVC in which there exists a basic architectural element capable of transitioning between two modes of behavior.  We call this element a "functional syllable unit" (FSU).  The modes are: "quiescence", in which the excitatory cells are silent above threshold, and "active", in which the excitatory cells are permitted to activate in a series, and where the activity is sustained via WLC.  Each mode can occur over a distinct range of values of the synaptic coupling strengths among the interneurons in the FSU.  We attribute a transition between modes to a neuromodulatory mechanism that is capable of altering those coupling strengths.  If a full song is taken to be comprised of a population of these FSUs, then we can illustrate how the observed population activity of both excitatory and inhibitory neurons during song - and during quiescence - in HVC can be reproduced.  That is: our model reproduces not only series activity of excitatory cells, but - more broadly - the behavior of both excitatory and inhibitory cells both during quiescence and during song, at the population level.  

In the next section we shall describe this proposed core HVC element, using Hodgkin-Huxley neurons with calcium dynamics.  The individual $HVC_I$ neurons have an experimental biophysical basis from the dissertation work of Daou et al. (2013) and subsequent experiments by Daou in the Margoliash laboratory at the University of Chicago (private communication); (Breen et al. 2016).  The model $HVC_{RA}$ neuron is based on experimentally observed currents (Daou et al. 2013; Long, Jin, \& Fee 2010) and is being tested further using protocols of Kadakia et al. (2016).

The remainder of the paper is structured as follows.
\begin{itemize}
  \item \textit{Constituents of a Functional Syllable  Unit} lays out the model neurons and synapses used in the functional architecture.
  \item In \textit{Results}, numerical results from the Hodgkin-Huxley based circuit show how the connectivity can give rise to two distinct modes of dynamics.  Furthermore, there exist narrow "transition regions" between these modes, within which lie additional modes that may occur on rare occasion; these modes should be identifiable in the laboratory.  
  \item In \textit{Building Complete Songs} we offer an example of creating a species-specific song using these FSUs as building blocks, thereby demonstrating how the basic qualitative population activity can be reproduced by our model. 
  \item In \textit{Discussion}, we examine winnerless competition as a theoretical framework for the competitive inhibitory neuron dynamics, and our suggestion that neuromodulation is the biophysical switch sculpting the HVC interneuron activity.  Then we turn to ways in which the model predictions can be tested experimentally.  
\end{itemize}
\end{multicols}

\section{\\A FUNCTIONAL SYLLABLE UNIT}
\begin{multicols}{2}
The central constituent of our model is a functional syllable unit (FSU).  We consider HVC to be comprised of numerous FSUs, each of which is comprised of an "inner" inhibitory loop that can assume one of two operational modes, depending on the synaptic coupling strengths among the inhibitory cells: (A): a state in which all inhibitory neurons fire continually, and (B): a state in which the inhibitory neurons are forced, via mutual inhibition, to fire in alternation.  We call these modes "quiescent" and "active", respectively.  

These two modes are captured in a Lotka-Volterra-like system, which expresses the quiescent mode when the inhibitory coupling strengths are weak; above some threshold value of coupling strengths, the circuit makes a bifurcation to the active mode.  This phenomenon of mode switching was formalized in another neural context as "winnerless competition" by Rabinovich et al. (2001). 

The inhibitory loop of $HVC_I$ neurons is connected to
\begin{figure}[H] 
\centering
  \includegraphics[width=70mm,valign=t]{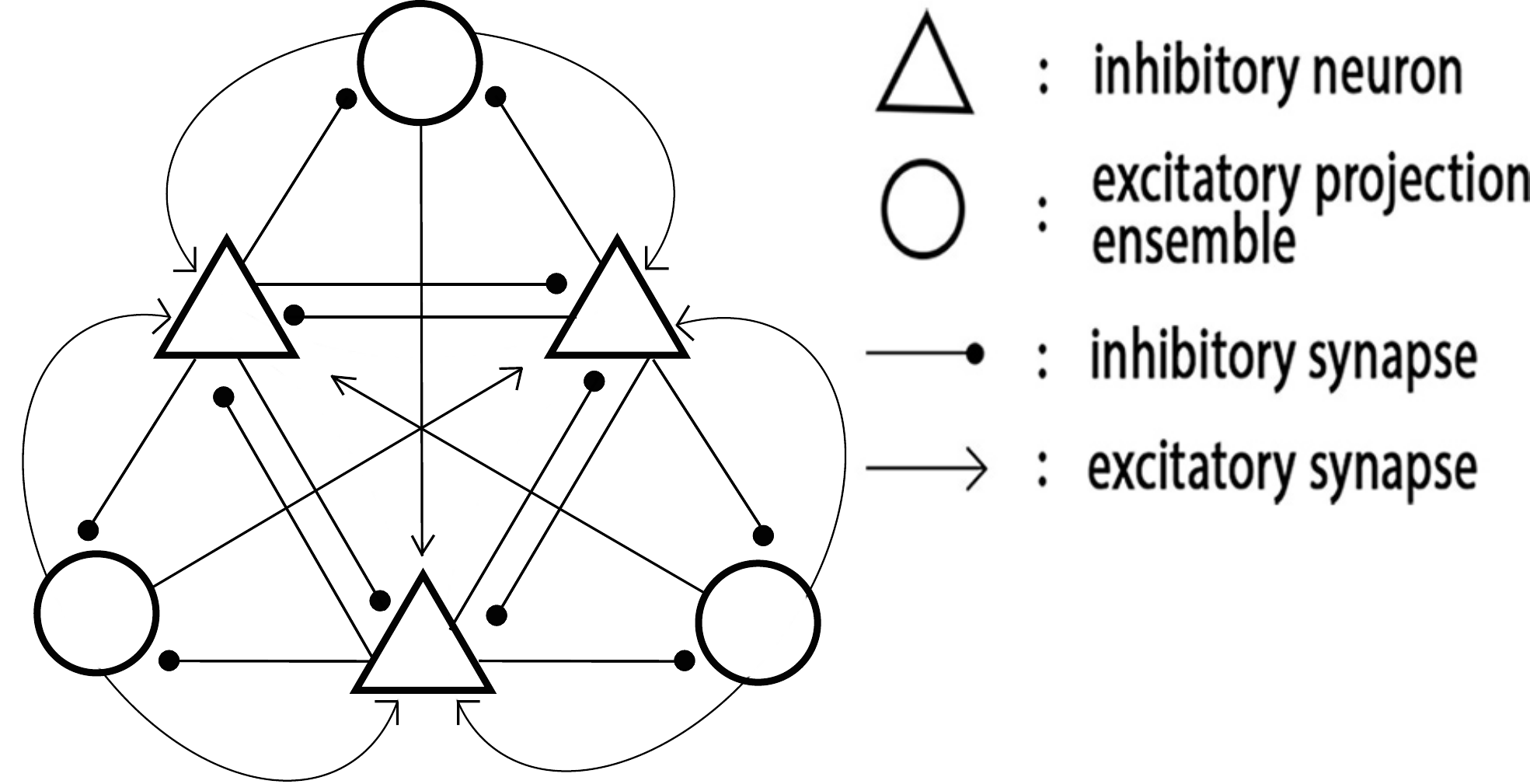}
  \caption{\textbf{A functional syllable unit (FSU)}, comprised of three $HVC_I$ neurons and three ensembles of $HVC_{RA}$ neurons.  Triangles and circles represent the former and latter populations, respectively.  Filled-circle- and arrow-headed lines represent inhibitory and excitatory functional connections, respectively.}
\end{figure}
\noindent
an "outer" loop of excitatory $HVC_{RA}$ projection neurons as shown in Figure 1.  Each triangle represents one inhibitory neuron and each circle represents an ensemble of $HVC_{RA}$ neurons.  The filled-circle- and arrow-headed lines represent inhibitory and excitatory functional synaptic connections, respectively.   

With this connectivity, we can make the following associations between behavior and FSU activity.  When the bird is not singing, all FSUs in HVC sit in a background of excitation sufficient for quiescence: all $HVC_I$ cells fire continually, thereby - via the wiring of Figure 1 - silencing all $HVC_{RA}$ neurons.  The coupling strengths in this FSU then are increased beyond the threshold required for a bifurcation to active mode.  This switching might be effected via an injection of an inhibitory neurotransmitter such as GABA, the primary type of inhibitory neurotransmitter in HVC (Dutar \& Perkel 1998; Schmidt \& Perkel 1998), in the vicinity of a particular FSU.  The $HVC_I$ neurons within that particular FSU now are forced into a winnerless competition, and they fire sequentially.  The wiring of Figure 1 then effects sequential firings of the $HVC_{RA}$ neurons.  To be clear: during its active state, an FSU represents the playing of one syllable, where - in the example architecture of Figure 1 - three $HVC_{RA}$ ensembles fire in sequence.  The syllable can play until the elevated inhibition drops below some threshold, at which point the FSU rapidly returns to quiescence.  

To model the full song of a specific species, we consider HVC to be a \textit{collection} of such pattern-generating structures.  While "Syllable A" is presumed to be initiated by an incentive to vocalize, we attribute subsequent syllables to a feedback loop, most likely from the motor area (e.g. Vallentin \& Long 2015), in which a series of FSUs in HVC is targeted by a succession of inhibitory neurotransmitter injections.  Such a loop was proposed by Gibb, Gentner, \& Abarbanel (2009b), who suggested the nucleus Uva as a possible source of the feedback.    
\end{multicols}

\section{\\CONSTITUENTS OF A FUNCTIONAL SYLLABLE UNIT}
\begin{multicols}{2}
\subsection{Neurons}
To build our model, we consider the $HVC_I$ and $HVC_{RA}$ populations.  HVC neurons projecting to Area X and the AFP are not considered.  We focus on the $HVC_{RA}$ and $HVC_I$ populations because they have been clearly identified as playing fundamental roles in song generation in the adult.

We consider the neurons to be Hodgkin-Huxley-type neurons all with sodium, potassium, and leak currents.  The specific inhibitory model we take from Breen et al. (2016), and we base our excitatory model on Kadakia et al. (2016).  Both of these models were constructed in light of recent electrophysiological data on calcium channels in these cells (Daou et al. 2013).  The $HVC_I$ neurons have T-type calcium and hyperpolarization-activated currents; the $HVC_{RA}$ neurons have L-type calcium channels and a potassium channel activated by increased intracellular calcium concentrations (hereafter the K/Ca channel).  In addition, the $HVC_{RA}$ neuron is treated as a two-compartment structure consisting of soma and dendrite.  The time evolution of the membrane potential V(t) of the neurons is expressed as follows. \\

\noindent
\textit{Inhibitory interneuron:}
\begin{align*} 
  C\diff{V_{i}(t)}{t} &= I_{L,i}(t) + I_{Na,i}(t) + I_{K,i}(t) + I_{CaT,i} + I_H 
\end{align*}
\begin{align}
        &+ \sum_{j \neq i}I_{syn,ij}(t) + I_{background} + noise(t)
\end{align}
\noindent
\textit{Excitatory projection neuron - somatic compartment:}
\begin{align*} 
  C\diff{V_{s,i}(t)}{t} &= I_{L,i}(t) + I_{Na,i}(t) + I_{K,i}(t) 
\end{align*}
\begin{align*}  
  &+ g_{SD}(V_d(t),V_s(t)) + I_{background} + noise(t)
\end{align*}
\setlength{\tabcolsep}{1pt}
\begin{table}[H]
\small
\centering
\begin{tabular}{ l c c|c c c c } \toprule
 \textit{Quantity} & Value &   & \textit{Quantity} & Value &   \\\midrule 
 \textit{$g_{L,0}$} & 0.00303 & $\mu$S & \textit{$g_{L,1}$} & 0.00302 & $\mu$S \\
 \textit{$g_{L,2}$} & 0.00299 & $\mu$S & \textit{$g_{L,3}$} & 0.00301 & $\mu$S \\
 \textit{$g_{L,4}$} & 0.00298 & $\mu$S & \textit{$g_{L,5}$} & 0.00297 & $\mu$S \\ 
 \textit{$E_{L,0}$} & -60.0 & mV & \textit{$E_{L,1}$} & -59.96 & mV \\
 \textit{$E_{L,2}$} & -59.94 & mV & \textit{$E_{L,3}$} & -80.0 & mV \\
 \textit{$E_{L,4}$} & -80.05 & mV & \textit{$E_{L,5}$} & -79.95 & mV \\
 \textit{$g_{H,1}$} & $2.0^{-3}$ & $\mu$S & \textit{$g_{H,2}$} & $1.99^{-3}$ & $\mu$S \\
 \textit{$g_{H,3}$} & $2.01^{-3}$ & $\mu$S & & & \\\midrule  
 \textit{$g_{Na}$} & 1.2 & $\mu$S & \textit{$g_{K}$} & 0.2 & $\mu$S \\
 \textit{$E_{Na}$} & 50.0 & mV & \textit{$E_{K}$} & -77.0 & mV \\
 \textit{$\theta_{m}$} & -40.0 & mV & \textit{$\theta_h$} & -60.0 & mV \\
 \textit{$\sigma_{m}$} & 16.0 & mV & \textit{$\sigma_h$} & -16.0 & mV\\
 \textit{$t_{0,m}$} & 0.1 & ms & \textit{$t_{0,h}$} & 1.0 & ms\\
 \textit{$t_{1,m}$} & 0.4 & ms & \textit{$t_{1,h}$} & 7.0 & ms\\ 
 \textit{C} & 0.01 & $\mu$F & \textit{$\theta_{n}$} & -55.0 & mV\\
 \textit{$g_{SD}$} &  0.05 & nS & \textit{$\sigma_{n}$} & 25.0 & mV\\
   &   &   &  \textit{$t_{0,n}$} & 1.0 & ms\\ 
  &   &  &  \textit{$t_{1,n}$} & 5.0 & ms\\\midrule
 \textit{$E_{H}$} & -40.0 & mV & \textit{$\theta_{H}$} & -60.0 & mV \\
 \textit{$\sigma_{0,H}$} & -11.0 & mV & \textit{$\sigma_{1,H}$} & 21.0 & mV \\
 \textit{$t_{0,H}$} & 0.1 & ms & \textit{$t_{1,H}$} & 193.5 & ms\\\bottomrule
\end{tabular}\\
\caption{\textbf{Parameter values for neuronal Na, K, L, and H currents.}  \textit{Top}: Neurons are distinguished via slightly different values of $g_L$ and $E_L$, where $g_{L,0}$ denotes the maximum leak conductance of cell zero (out of five).  $HVC_I$ neurons are further distinguished by different values of $g_H$.  \textit{Middle}: Values for Na, K, and L current kinetics and other cellular properties.  \textit{Bottom}: Values for H current kinetics.  Units: mV are milli-Volts; ms are milli-seconds; $\mu$F are micro-Farads; $\mu$S are micro-Siemens.} 
\end{table}
\setlength{\tabcolsep}{1pt}
\begin{table}[H]
\small
\centering
\begin{tabular}{ l c c|c c c c } \toprule
 \textit{Quantity} & Value &   & \textit{Quantity} & Value &   \\\midrule 
 \textit{$g_{CaT,0}$} & $1.0^{-4}$ & $\mu$S & \textit{$g_{CaT,1}$} & $1.01^{-4}$ & $\mu$S \\
 \textit{$g_{CaT,2}$} & $1.01^{-4}$ & $\mu$S & \textit{$g_{CaL,3}$} & 0.00301 & $\mu$S \\
 \textit{$g_{CaL,4}$} & 0.00298 & $\mu$S & \textit{$g_{CaL,5}$} & 0.00297 & $\mu$S \\ 
 \textit{$g_{KCa,3}$} & -60.0 & mV & \textit{$g_{KCa,4}$} & -59.96 & mV \\
 \textit{$g_{KCa,5}$} & -59.94 & mV & &  & \\\midrule  
 \textit{$\theta_{a}$} & -70.0 & mV & \textit{$\theta_b$} & -65.0 & mV \\
 \textit{$\sigma_{a}$} & 10.0 & mV & \textit{$\sigma_b$} & -10.0 & mV\\
 \textit{$t_{0,a}$} & 0.1 & ms & \textit{$t_{0,b}$} & 1.0 & ms\\
 \textit{$t_{1,a}$} & 0.2 & ms & \textit{$t_{1,b}$} & 5.0 & ms\\
 \textit{C} & 0.01 & $\mu$F & \textit{$\theta_{q}$} & -40.0 & mV\\
 \textit{$Ca_{ext}$} & 2500. & $\mu$M & \textit{$\sigma_{q}$} & 10.0 & mV\\
 \textit{$k_s$} & 2.5 & $\mu$M & \textit{$t_{0,q}$} & 1.0 & ms\\ 
 \textit{$\phi$} & 0.06 & $\mu$M/ms/nA & \textit{$t_{1,q}$} & 0.0 & ms\\
 \textit{$\tau_{CA}$} & 10. & ms & \textit{$Ca_{0}$} & 0.2 & $\mu$M \\\bottomrule
\end{tabular}\\
\caption{\textbf{Parameter values for calcium dynamics}.  \textit{Top}: Neurons are distinguished via slightly different values of $g_{CaT}$ (for $HVC_I$ neurons) and $g_{CaL}$ and $g_{KCa}$ (for $HVC_{RA}$ neurons).  \textit{Bottom}: Values for CaT and CaL current kinetics and basic cellular properties.  Values were chosen based on electrophysiology of HVC neurons \textit{in vitro} (Daou et al. 2013).} 
\end{table}
\noindent
\textit{Excitatory projection neuron - dendritic compartment:}
\begin{align*} 
  C\diff{V_{d,i}(t)}{t} &= I_{CaL,i}(t) + I_{KCa,i}(t) + g_{SD}(V_s(t),V_d(t))    
\end{align*}
\noindent
where C is the membrane capacitance and noise(t) is a low-amplitude background noise term.  The $I_{syn}$ terms represent the synaptic input currents, $I_{background}$ is a DC current representing ambient background excitation, and the $g_{SD}$ terms couple the compartments.  The ion channel currents for the $i^{th}$ neuron are: 
\begin{align*} 
  I_{L,i}(t) &= g_{L}(E_{L} - V_i(t))\\ 
  I_{Na,i}(t) &= g_{Na,i} m(t)^3 h(t) (E_{Na} - V_i(t)) \\
  I_{K,i}(t) &= g_{K,i} n(t)^4 (E_K - V_i(t))\\
  I_{CaT,i}(t) &= g_{CaT,i} a(t)^3 b(t)^3 GHK(V_i(t),[Ca]_i(t))\\
  I_{CaL,i}(t) &= g_{CaL,i} q(t)^2 GHK(V_i(t),[Ca]_i(t))  \\
  I_{KCa,i} &= g_{KCa}\frac{[Ca]_i(t)^2}{[Ca]_i(t)^2 + k_s^2}(E_K - V_{d,i}(t))\\  
  I_{H,i}(t) &= g_{H}H(t)^2(E_{H} - V_i(t))\\
\end{align*}
\noindent
where "GHK($V_i$(t),$[Ca]_i(t)$)" is defined as:
\begin{align*} 
  GHK(V_i(t),[Ca]_i(t)) &= V_i(t)\frac{[Ca]_i(t) - Ca_{ext}e^{-2FV_i(t)/RT}}{e^{-2FV_i(t)/RT} - 1}.
\end{align*} 
\noindent
The parameters denoted "g" are the maximum conductances of each current; the parameters denoted "E" are the respective reversal potentials.  [Ca](t) is the intracellular $Ca^{2+}$ concentration as a function of time.  $Ca_{ext}$ is the extracellular concentration of $Ca^{2+}$ ions.  In the GHK current, F is the Faraday constant, R is the gas constant, and T is temperature, which we take as 37$^{\circ}$ C.  The gating variables $U_i(t)$ = [m(t), h(t), n(t), a(t), b(t), q(t), H(t)] satisfy:
\begin{align*} 
  \diff{U_i(t)}{t} &= (U_{\infty}(V_i(t)) - U_i(t))/\tau_{Ui}(V_i(t)); \\
  U_{\infty}(V_i) &= 0.5 [1 + \tanh((V_i - \theta_{U,i})/\sigma_{U,i})]\\
  \tau_{Ui}(V_i) &= t_{U0} + t_{U1}[1 - \tanh^2((V_i - \theta_{U,i})/\sigma_{U,i})]. 
\end{align*} 
\noindent
There is one exception for H(t): $H_{\infty}$ and $\tau_H$ take different values of $\sigma_{0,H}$ and $\sigma_{1,H}$ (see Table 1).  The calcium dynamics evolve as:
\begin{align*} 
  \diff{[Ca_i](t)}{t} &= \phi I_{CaX} + \frac{Ca_0 - [Ca_i](t)}{\tau_{Ca}}
\end{align*}
\noindent
where the "X" of the subscript "CaX" represents "T" (for T-type) and "L" (for L-type), for the inhibitory neuron and excitatory dendrite, respectively.  $Ca_0$ is the equilibrium concentration of calcium inside the cell, and $k_s$ is a Michaelis-Menten constant. 

We further distinguish the excitatory and inhibitory neurons via their respective resting potential $E_L$ and synaptic reversal potential $E_{syn,rev}$.  We render the neurons distinguishable via slightly different values of their leak maximum conductances $g_L$ and resting potentials $E_L$ (for all neurons), CaT and H current maximum conductances $g_{CaT}$ and $g_{H}$  (for the interneurons), and CaL and KCa current maximum conductances $g_{CaL}$ and $g_{KCa}$ (for the excitatory projection neurons).  Values for the Na, K, L, and H currents and basic cellular properties are listed in Table 1; values for the calcium dynamics are in Table 2.  

\subsection{Synapses}

For the synapse dynamics, we adopt the formalism of Destexhe \& Sejnowski (2001) and Destexhe et al. (1994) for electrically-delivered neurotransmitter pulses, with one alteration: we define the inhibitory synapse coupling strengths $g_{ij}$ not as fixed numbers but rather as functions of the maximum neurotransmitter concentration $T_{max}$ presented to a post-synaptic neuron.  Within this framework, $T_{max}$ itself is a function of some modulatory process that may be external to HVC.  Details of this formulation are presented in Part E of \textit{Results}. 
\begin{align} 
  I_{syn,ij} &= g_{ij}(T_{max}(t))s_{ij}(t)(E_{syn,i} - V_{i}(t))\\  
  \diff{s_{ij}(t)}{t} &= \alpha(T_{max}(t),V_j(t))[1 - s_{ij}(t)] - \beta s_{ij}(t)\\   
  \alpha(T_{max}(t),V_j(t)) &= \frac{T_{max}/T_0}{1 + \exp(-(V_{j}(t) - V_{P})/K_{P}).} 
\end{align}
\noindent
$I_{syn,ij}$ is the current seen by post-synaptic cell i as a result of input from pre-synaptic cell j.  $E_{syn,i}$ is the synaptic reversal potential of cell i, $V_i$(t) is the instantaneous membrane voltage of cell i, and $s_{ij}$(t) is the gating variable of the synapse entering (post-synaptic) cell i from (pre-synaptic) cell j.  $T_0$ has units of ms-mM so that $\alpha$($T_{max}$,V), the rate of gate opening, has units of 1/time; $\beta$, the rate of gate closing, also has units of 1/time.  $V_{j}$(t) is the pre-synaptic membrane voltage, and $V_{P}$ and $K_{P}$ are parameters governing the shape of the distribution of neurotransmitter rise and fall as it drives gating variables $s_{ij}$.  Parameter values for the synapse equations are given in Table 3.

We consider two broad classes of neurotransmitter: excitatory and inhibitory.  The maximum concentration of \textit{excitatory} neurotransmitter we take to have a constant value of 1.5 mM\footnote{Throughout this paper, we will adopt the custom of referring to concentration in moles.  The custom exists because of the many uncertainties involved in determining synapse volumes (see \textit{Discussion}).}; the maximum concentration of \textit{inhibitory} neurotransmitter ($T_{max,inh}$) we permit to vary.  We define the inhibitory-to-inhibitory coupling strengths ($g_{ij,inh-to-inh}$) so that they increase with $T_{max,inh}$.  $T_{max,inh}$ is some function of the activity of a cell that may be external to HVC.  We will discuss the selection of $T_{max,inh}$, and the dependence of $g_{ij}$ on that value, in Part E of \textit{Results}.  \setlength{\tabcolsep}{1pt}
\begin{table}[H]
\small
\centering
\begin{tabular}{ l c c|c c c c } \toprule
 \textit{Quantity} & Value  &   & \textit{Quantity} & Value &   \\\midrule 
 \textit{$E_{syn,inh}$} & -80.0 & mV & \textit{$\beta_{inh}$} & 0.18 & $ms^{-1}$ \\
 \textit{$E_{syn,exc}$} & 0.0 & mV & \textit{$\beta_{exc}$} & 0.38 & $ms^{-1}$\\
 \textit{$V_{p}$} & 2.0 & mV & $T_0$ & 1 & ms-mM  \\
 \textit{$K_{p}$} & 5.0 & mV \\\bottomrule
\end{tabular}\\
\caption{\textbf{Parameter values for synapses.}} 
\end{table}
\end{multicols}

\section{\\RESULTS}
\begin{multicols}{2}
In this section we illustrate, via the time course of membrane voltages of an FSU's constituent neurons, how an FSU functions dynamically.  The steps are as follows.
\begin{itemize}
  \item First we will demonstrate that the excitatory neurons of an FSU will fire when given a low background excitation.  
  \item Second, we will show that imposing sufficient inhibition upon these excitatory neurons leads to a quiescent FSU, for a low range of inhibitory-to-inhibitory coupling strengths.  
  \item Third, we will show that a higher range of these coupling strengths, combined with a higher value of $T_{max}$, can effect a regime in which the inhibitory neurons alternate their firing patterns; this scenario represents an active FSU.  
  \item Fourth, by exploring the behavior of an FSU over ranges of both $T_{max}$ and the synapse coupling strengths $g_{ij}$, we will demonstrate: a) the quiescent and active modes are robust to small variations in these parameter values; b) there exist additional transient modes of behavior, which one might expect to occasionally encounter in the laboratory.  
  \item Finally, we will note that connections from the excitatory to inhibitory cells are critical for the stable propagation of a series in active mode.  
\end{itemize}
For all voltage time series shown in this paper, the dynamical model was written in Python, and the equations of motion were integrated using Python's adaptive fourth-order Runge-Kutta "odeINT" with a step of 0.1 millisecond.

\subsection{Essential architecture of an FSU}

Our FSU is constructed as depicted in Figure 1.  The essential features of the connectivity are these:
\begin{itemize}
  \item There is all-to-all connectivity among the $HVC_I$ neurons;
  \item Each $HVC_I$ neuron synapses to neurons in two out of three of the $HVC_{RA}$ ensembles as shown, where no two $HVC_I$ neurons "omit" the same $HVC_{RA}$ ensemble; 
  \item Neurons within each $HVC_{RA}$ ensemble synapse to all three $HVC_I$ neurons as shown.
\end{itemize}
With this connectivity, each $HVC_{RA}$ ensemble connects disynaptically, via an interneuron, to both of the other $HVC_{RA}$ ensembles.  There are no monosynaptic connections between $HVC_{RA}$ ensembles in an FSU.

We leave unspecified: i) the rate of excitatory-to-excitatory connections within an ensemble; ii) the number of neurons within each ensemble to which one $HVC_I$ neuron projects; iii) the number of neurons within an ensemble that project to a particular $HVC_I$ neuron.  The number of $HVC_{RA}$ neurons per ensemble may be taken to be eight, for agreement with the observed \textasciitilde 8:1 ratio of $HVC_{RA}$ to $HVC_I$ neurons in the nucleus; however, our model does not require any particular value.

\subsection{$HVC_{RA}$ neurons spike in absence of inhibition}
We first sought to simulate an environment in which $HVC_{RA}$ neurons will, in the absence of inhibition, fire above the threshold required to generate an action potential - the environment that is indicated by the results of KVL15.  To this end, we gave the three $HVC_{RA}$ ensembles of Figure 1 an injected current of 0.3 nA with a random variation of 3 $\%$.  Figure 2 shows the voltage traces of three $HVC_{RA}$ cells, one in each ensemble.  All projection cells spike. 
\begin{figure}[H]
\centering
  \includegraphics[width=70mm]{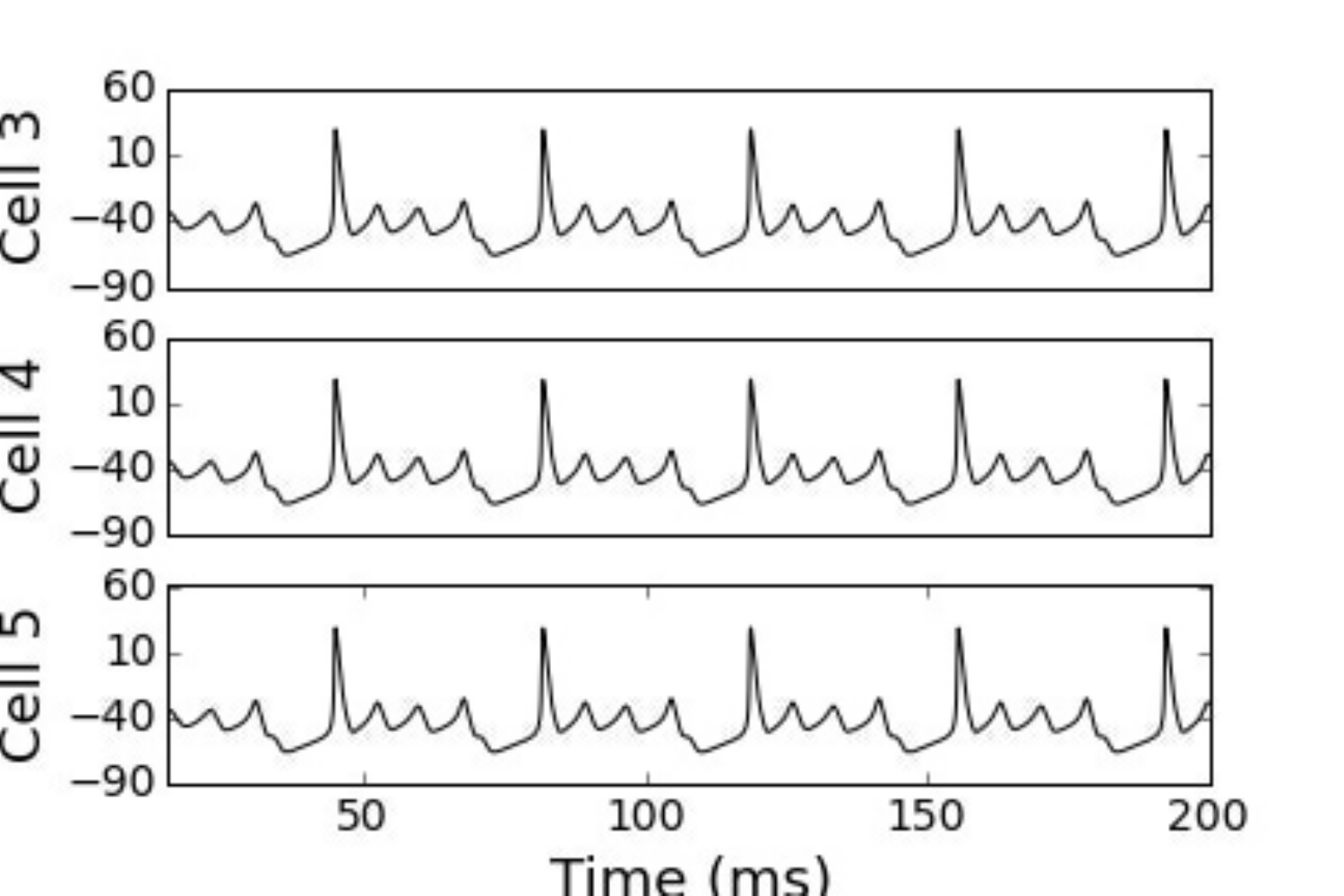}\\
  \caption{\textbf{Voltage traces of three $HVC_{RA}$ projection neurons}, one in each of the three ensembles.  They fire when given a low background current, in the absence of inhibition.  The background current is 0.3 nA with a random 3 $\%$ variation.}
\end{figure}
\noindent
\subsection{A quiescent FSU}
Now we involve the interneurons.  First we connect the interneurons all-to-all and set these synapse strengths at values near 0.01 $\mu$S (see Table 4), and with the value of $T_{max}$ at an initial low value of 0.5 mM.  With these choices, the $HVC_I$ cells fire continually.  

Next we permit each $HVC_I$ neuron to synapse directly to two of the three $HVC_{RA}$ ensembles, where no two $HVC_I$ neurons "omit" the same $HVC_{RA}$ ensemble.  We set these inhibitory-to-excitatory connection strengths to be of order 1.0 $\mu$S.  With these choices, the inhibition is sufficiently strong to overpower the background stimulation of 0.3 nA and silence all the $HVC_{RA}$ ensembles.  That is: each $HVC_I$ neuron actively suppresses two of the $HVC_{RA}$ ensembles. 
\setlength{\tabcolsep}{1.5pt}
\begin{table}[H]
\small
\centering
\begin{tabular}{ l c c c c c c } \toprule
 \textit{Cell} & \textit{0} & \textit{1} & \textit{2} & \textit{3} & \textit{4} & \textit{5} \\\midrule 
 \textit{0} & 0.0 & 0.011 & 0.011 & 1.11 & 1.1 & 1.11 \\
 \textit{1} & 0.011 & 0.0 & 0.01 & 1.11 & 1.1 & 1.1  \\
 \textit{2} & 0.011 & 0.011 & 0.0 & 1.11 & 1.1 & 1.1 \\
 \textit{3} & 1.1 & 1.11 & 0.0 & 0.0 & 0.0 & 0.0\\
 \textit{4} & 0.0 & 1.11 & 1.1 & 0.0 & 0.0 & 0.0 \\
 \textit{5} & 1.11 & 0.0 & 1.11 & 0.0 & 0.0 & 0.0 \\\bottomrule
\end{tabular}\\
\caption{\textbf{Synapse strengths $g_{ij}$ for the voltage traces in Figure 3 for a quiescent FSU}.  Units are micro-Siemens ($\mu$S).  The value of 0.011 $\mu$S in [row 0, column 1] corresponds to $g_{01}$: the synapse entering Cell 0 from Cell 1.  Here, the inhibitory-to-inhibitory coupling strengths are of order 0.01 $\mu$S (they must be below \textasciitilde 0.5 $\mu$S for continual firing of the $HVC_I$ neurons to occur); the inhibitory-to-excitatory coupling strengths must be above \textasciitilde 1 $\mu$S to overpower the background excitation and silence the $HVC_{RA}$ neurons; the excitatory-to-inhibitory connections here are of order 1 $\mu$S, but they need not be any particular value for continual activity of the $HVC_I$ neurons in the FSU to occur.}
\end{table}
Figure 3 illustrates this result.  The voltage traces of the $HVC_I$ cells (top) shows that they are firing continually.  As each $HVC_I$ neuron projects to two $HVC_{RA}$ ensembles, the three previously uninhibited $HVC_{RA}$ ensembles are suppressed.  The corresponding schematic is shown at bottom, where black- and white-filled symbols represent neural activity above and below threshold, respectively.  This regime we call quiescence.  

\subsection{An active FSU}
To transition from quiescent to active mode, our FSU requires two modifications:
\begin{itemize}
  \item the $HVC_I$-to-$HVC_I$ coupling strengths must increase by roughly two orders of magnitude: from \textasciitilde 0.01 to 2.0 $\mu$S (see Table 5);
  \item $T_{max}$ must increase by a factor of roughly three: from \textasciitilde 0.5 to 1.8 mM.  
\end{itemize}
\noindent
This second requirement implies the direct correlation between synapse coupling strengths and neurotransmitter concentration that we have embodied in Equations 7 and 8. 

With these adjustments, and with the same background current, the $HVC_I$ cells now alternate their spiking activity.  Given the wiring, the activity of the $HVC_{RA}$ ensembles also alternates.  The alternations are a particular repeating series.  Figure 4 illustrates this activity in a three-frame "movie".  

At left in Figure 4 are three pairs of voltage traces.  Each pair represents the activity of one $HVC_I$ neuron and one $HVC_{RA}$ ensemble, where the $HVC_{RA}$ ensemble in each pair is the one $HVC_{RA}$ ensemble in the FSU to which that particular $HVC_I$ neuron does \textit{not} project.  At right are the corresponding schematics, where each pair is sequentially highlighted by a specific color.  Cells in each pair may fire simultaneously.  By this wiring, series activity of the $HVC_I$ neurons effects a series of activity of the $HVC_{RA}$ ensembles. 

First, the "green pair" fires (top row).  Here, the green $HVC_I$ neuron has activated and is able to suppress the other two $HVC_I$ neurons, which are colored in white in the schematic to indicate that they are currently inactive above threshold.  The active (green) $HVC_I$ neuron projects to two of the three $HVC_{RA}$ ensembles; these two are colored in white in the schematic to indicate that they are not active above threshold.  The third $HVC_{RA}$ ensemble, to which the active $HVC_I$ neuron does \textit{not} project, is the only uninhibited ensemble.  In the presence of the background excitation, this $HVC_{RA}$ ensemble bursts (starting around $t=250 ms$); hence it is also colored green.  

\begin{figure}[H]
  \includegraphics[width=90mm]{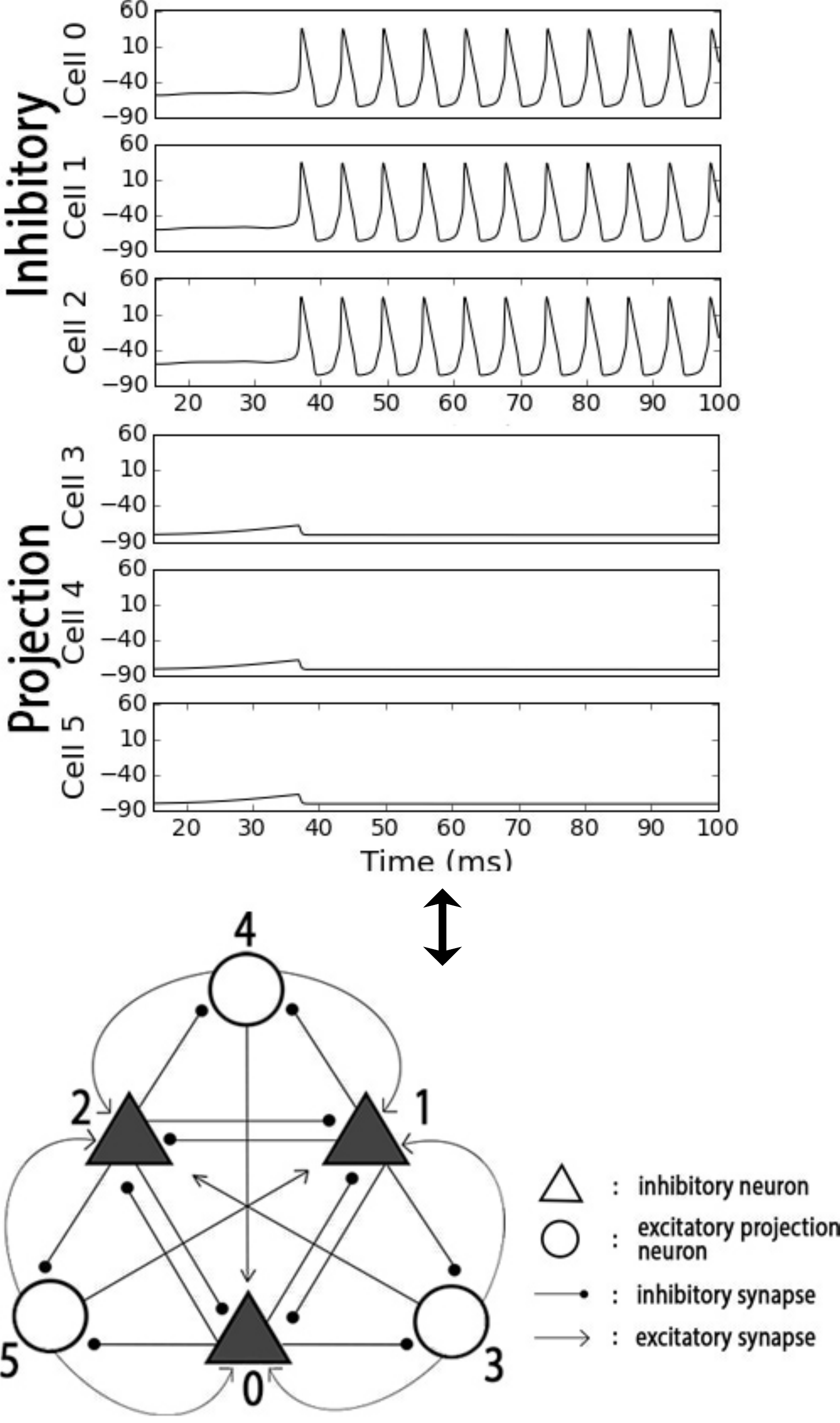}\\
  \caption{\textbf{A quiescent FSU}.  Voltage traces of the three $HVC_I$ neurons (\textit{top}) and three $HVC_{RA}$ neurons each representing one ensemble (\textit{middle}), all within a quiescent FSU, where the inhibitory-to-inhibitory coupling strengths (Table 4) are sufficiently low to permit the $HVC_I$ neurons to fire continually.  $T_{max}$ $=$ 0.5 mM.   \textit{Bottom}: The corresponding schematic, where triangles (cells 0, 1, and 2) and circles (cells 3, 4, and 5) represent $HVC_I$ and $HVC_{RA}$ neurons, respectively.  The cell numbers on the schematic correspond to the numbering of the voltage traces.  Black and white shapes indicate activity above and below threshold, respectively.  Each $HVC_I$ projects to two $HVC_{RA}$ neurons.  When all three $HVC_I$ neurons are active simultaneously, the three $HVC_{RA}$ ensembles are suppressed.}   
\end{figure}

\end{multicols}
\begin{figure}[H]
  \centering
  \includegraphics[width=120mm]{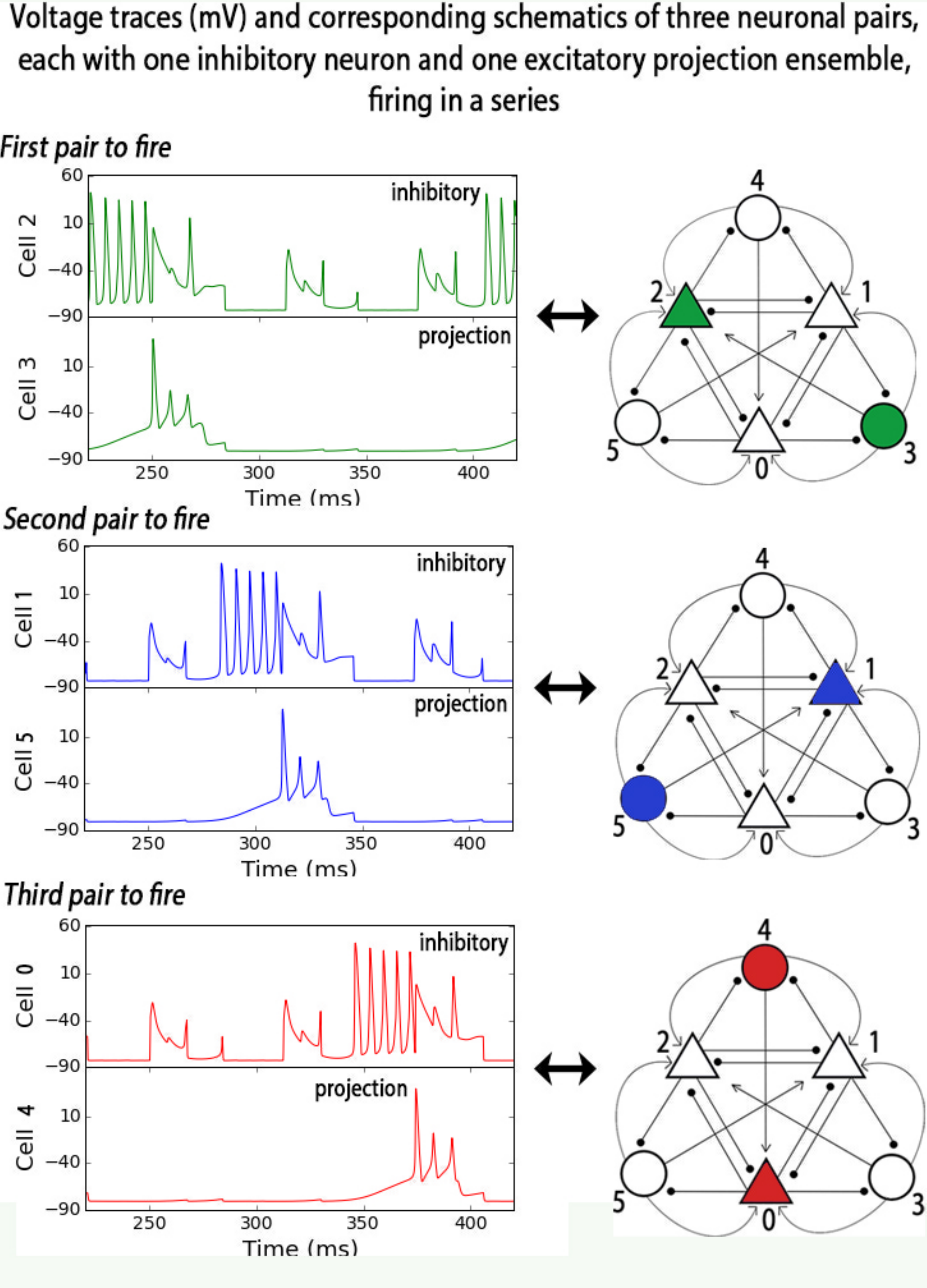}\\
  \caption{\textbf{An active FSU, represented in a three-frame "movie".}  Here, the inhibitory-to-inhibitory coupling strengths are roughly 2 $\mu$S Table 5) and $T_{max}$ $=$ 1.8 mM, so that the inhibitory neurons may engage in competitive dynamics.  \textit{Left}: Three "pairs" of voltage traces, each representing one $HVC_I$ neuron (top panel in each pair) and the one $HVC_{RA}$ ensemble to which that particular $HVC_I$ neuron does \textit{not} directly project (bottom panel in each pair).  Cells designated as "pairs" may fire simultaneously.  Thus, series activity of the $HVC_I$ neurons effects a series of activity of the $HVC_{RA}$ ensembles.  \textit{Right}: The corresponding schematic of each pair, where a currently-active pair is highlighted by a specific color.  The activity proceeds clockwise, beginning with the "green pair" (\textit{top row}), followed by the "blue pair" second (\textit{middle}), and "red pair" third (\textit{bottom}).  The numbering of cells on the schematic corresponds to the numbering of the voltage traces.}  
\end{figure}
\begin{multicols}{2}

Next, activity in the "inner loop" of inhibitory neurons proceeds clockwise, to the "blue" $HVC_I$ neuron.  An analogous situation to the above is now established, wherein the active (blue) $HVC_I$ neuron suppresses the other two $HVC_I$ neurons and two of the $HVC_{RA}$ ensembles, and the one unsuppressed $HVC_{RA}$ ensemble (also blue) may burst (at t \textasciitilde 315 ms).  Finally, in the third row, the third "red pair" bursts at t \textasciitilde 375 ms. 

This active FSU represents one syllable of song during which three ensembles of $HVC_{RA}$ neurons fire in a sequence.  Note again that to effect this series, along with the increase in the inhibitory coupling strengths, an increase in $T_{max}$ from 0.5 to 1.8 mM was required.  

Figure 5 demonstrates that this particular series repeats, as long as there are elevated values of $g_{ij}$ and $T_{max}$.  The six voltage traces are shown, now over three times the duration of Figure 4.  The panels are arranged so that the voltage traces of the three inhibitory and excitatory neurons are shown at top and bottom, respectively.  This organization is intended to facilitate recognition that the inhibitory (excitatory) cells all possess roughly identical waveforms with phase offsets. 

Table 5 lists the synapse strengths for this active FSU, showing that the strengths of the inhibitory-to-inhibitory connections are roughly two orders of magnitude greater than their quiescence values.  We note one concern regarding the relative weights of these strengths in active mode: they had to be close, to one part in 20, for the series WLC to occur.  This point is addressed in \textit{Discussion}.
\begin{figure}[H]
  \centering
  \includegraphics[width=70mm]{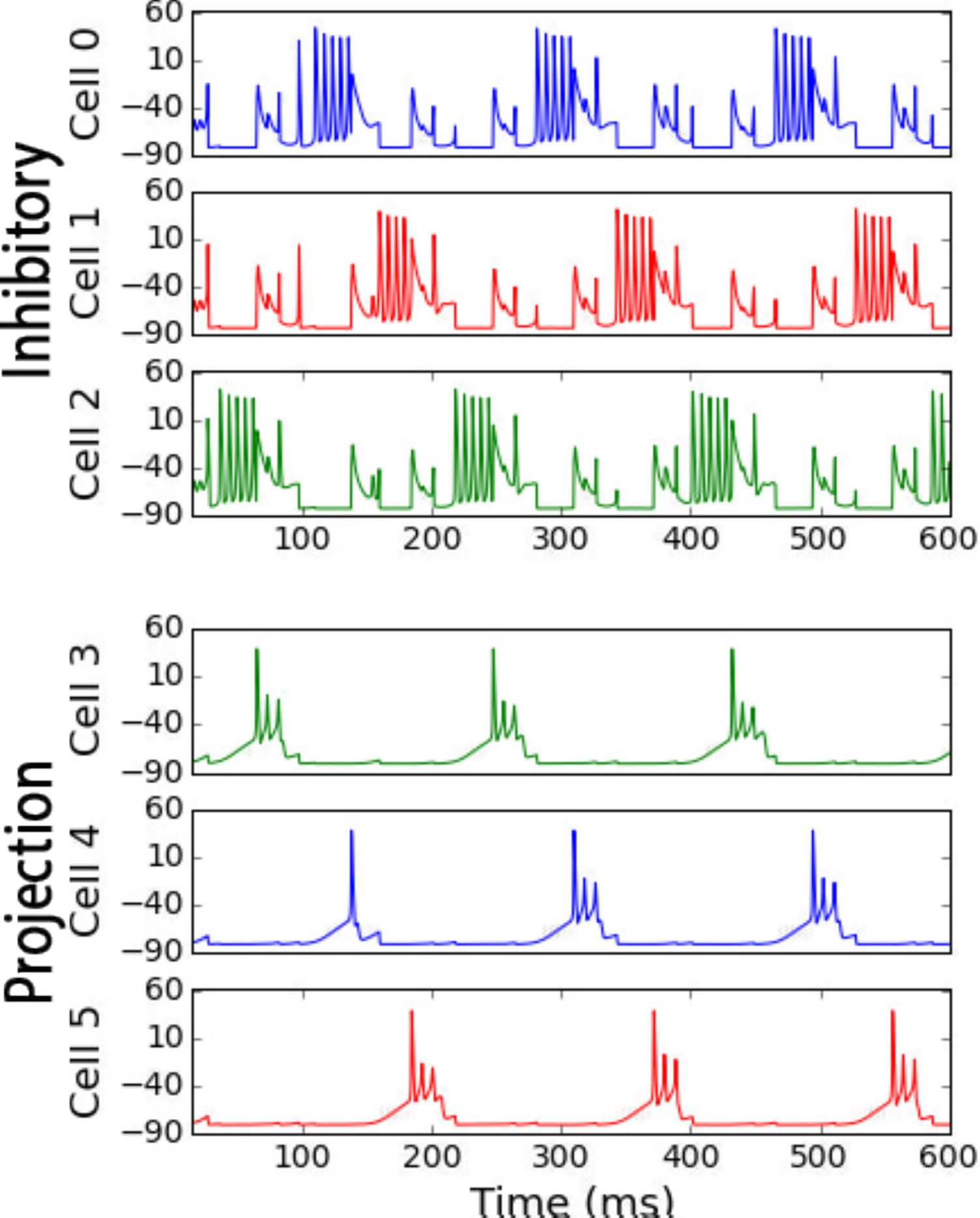}\\
  \caption{\textbf{Three rotations of the series activity represented in Figure 4}, where the three inhibitory cells and excitatory ensembles are each grouped together at top and bottom, respectively.  Color coding is as defined in Figure 4.}
\end{figure}
\subsection{Behavior of an FSU in terms of $T_{max}$ and $g_{ij}$}
We have noted that the activity of an FSU can be altered by effecting changes in the couplings among $HVC_I$ neurons and the maximum neurotransmitter concentration $T_{max}$ in the synaptic gating variable driving function $\alpha$($T_{max}$,V), where we use "$T_{max}$" to refer to the concentration of neurotransmitter governing inhibitory activity\footnote{As noted earlier, in this model we assume that the maximum neurotransmitter concentration governing the \textit{excitatory} connections is a constant.  It is, of course, possible that a modulation of the excitatory synaptic connections is also occurring.  This consideration will enter into the construction of a more generalized model.}.  
\setlength{\tabcolsep}{1.5pt}
\begin{table}[H]
\small
\centering
\begin{tabular}{ l c c c c c c } \toprule
 \textit{Cell} & \textit{0} & \textit{1} & \textit{2} & \textit{3} & \textit{4} & \textit{5} \\\midrule 
 \textit{0} & 0.0 & 2.1 & 2.0 & 1.11 & 1.1 & 1.11 \\
 \textit{1} & 2.1 & 0.0 & 2.1 & 1.11 & 1.1 & 1.1  \\
 \textit{2} & 2.0 & 2.1 & 0.0 & 1.11 & 1.1 & 1.1 \\
 \textit{3} & 1.1 & 1.11 & 0.0 & 0.0 & 0.0 & 0.0\\
 \textit{4} & 0.0 & 1.11 & 1.1 & 0.0 & 0.0 & 0.0 \\
 \textit{5} & 1.11 & 0.0 & 1.11 & 0.0 & 0.0 & 0.0 \\\bottomrule
\end{tabular}\\
\caption{\textbf{Synapse strengths $g_{ij}$ for the voltage traces in Figure 4 of an active FSU}.  Units are micro-Siemens ($\mu$S).  Notation: the value of 2.1 $\mu$S in [row 0, column 1] corresponds to $g_{01}$: the synapse entering Cell 0 from Cell 1.  Here, the inhibitory-to-inhibitory coupling strengths must reach a threshold roughly 2 $\mu$S for a stable series of bursting activity of $HVC_{RA}$ neurons to occur.   The inhibitory-to-excitatory strengths may remain of order 1 $\mu$S; the excitatory-to-inhibitory strengths must reach a value of order 1 $\mu$S for the series of $HVC_I$ to occur.}
\end{table}
\noindent
In this framework, $T_{max}$ itself is a function of some unspecified mechanism that may be external to HVC (see \textit{Discussion}).  To characterize the range and stability of an FSU's behavioral modes, we examined the activity of the FSU's cells over a broad set of values for $T_{max}$ and the inhibitory-to-inhibitory coupling strengths $g_{ij}$.  Figure 6 shows these results.  It depicts FSU behavior over the ranges $T_{max}$:[0.5,10.5] mM and $g_{ij}$:[0.01,10.5] nS, where we characterize "behavior" in terms of the $HVC_{RA}$ ensemble activity.  The symbols that represent this activity are defined to the right of the plot.  

We note three items regarding Figure 6.  First, the quiescent (solid blue circles) and active WLC bursting (green plus signs) modes are the most common behaviors, and they are robust to small variations in $T_{max}$ and $g_{ij}$: both span wide ranges of these parameter values.  In fact, the top row, at $T_{max}$ $=$ 10.5, looks the same up to a $T_{max}$ value of 50 mM.   Second, while the transition regions between quiescent and active mode are narrow compared to the permitted parameter ranges for these two dominant modes, the regions do possess a finite area - within which lie additional modes of behavior.  We will return to these additional modes later.  Third, we have drawn a black line on the plot.  This line represents a "path" through ($T_{max}$,$g_{ij}$) space, parameterized by time, that is capable of representing an FSU's transition between quiescent and active mode.  We shall now discuss the significance of such a path and the method used to construct the particular path depicted in Figure 6.

Using the known characteristics of a single song syllable as our guide, we sought to construct an equation for $T_{max}$(t) and a relation between $T_{max}$(t) and $g_{ij}$(t) that might permit an FSU to transition seamlessly between the quiescent and active states described earlier in this paper.  Any such path in ($T_{max}$,$g_{ij}$) space will produce a particular waveform of voltage behavior in both the $HVC_I$ and $HVC_{RA}$ neurons in an FSU, depending on which modes (the symbols in Figure 6) are visited over the course of that path, and the rate at which they are traversed. 

A detailed exploration of the classes of paths and their manifestations as voltage activity in HVC is beyond the scope of this paper; however, it will constitute a major topic of future investigations of FSUs.  In this paper we explore a single path in this space, which, again, we selected by requiring that it be capable of effecting the quiescent and active behavior described earlier.  

We began our path design by first asking that $T_{max}$(t) sit at some minimum value until the FSU receives a neurotransmitter injection at time t$=$0.  At t$=$0, $T_{max}$ should then rise rapidly with time.  $T_{max}$ reaches a maximum value at some time $t_1$, and then it instantly begins to decay more gradually.  This process is captured by the following three steps:
\begin{align}
  T_{max}(t) &= T_{max}^{lower bound}\hspace{1em} (t < 0)\\
  T_{max}(t) &= T_{max}^{lower bound} e^{t/\tau_r}\hspace{1em}  (0 < t < t_1)\\ 
  T_{max}(t) &= \vartheta T_{max}^{upper bound} e^{-t/\tau_f}\hspace{1em} (t > t_1),
\end{align}
\noindent
where $\tau_r$ and $\tau_f$ are constants dictating the rates of rise and fall, respectively, $t_1$ is the time of transition from rise to fall:
\begin{align*}
  t_1 = \tau_r\:  log(\frac{T_{max}^{upper bound}}{T_{max}^{lower bound}}),
\end{align*}
\noindent
and $\vartheta$ is a constant chosen for continuity at $t_1$:
\begin{align*}
  \vartheta = \frac{T_{max}^{lowerbound}}{T_{max}^{upperbound}}e^{(t_1/t_r + t_1/t_f)}.
\end{align*}
\noindent
$T_{max}^{lowerbound}$ and $T_{max}^{upperbound}$ are the lower and upper bounds of $T_{max}$, respectively.  We chose $T_{max}^{lowerbound}$ to be 0.5 mM so that when the stimulating neurotransmitter signal arrives, the FSU will be in quiescent mode.  We chose $T_{max}^{upperbound}$ to be 5.0 mM, a value that falls well within the permitted range for active behavior (Figure 6).  The rise and fall constants were chosen, in light of the relation between $T_{max}$ and $g_{ij}$ to be discussed below, to mimic the time course of a typical syllable of song.  The rise constant $\tau_r$ was taken to be 1 ms for a rapid transition from quiescent to active mode.  The fall constant $\tau_f$ was taken to be 4 ms so that active mode would be sustained for \textasciitilde 200 ms, the typical duration of a song syllable.  

To model the response of the synapse strengths $g_{ij}$ to $T_{max}$(t), we have taken the form:
\begin{align}
  g_{ij}(T_{max}) &= \mu T_{max}^\gamma,
\end{align}
\noindent
and we require that $g_{ij}$ rise no farther than a preselected value, which occurs when $T_{max}$ rises above a critical value of $T_{saturation}$.  We have imposed this requirement in order to achieve the rapid rise in the strengths $g_{ij}$ with $T_{max}$ while avoiding the (otherwise inevitable) exponential increase of $g_{ij}$ to values that are too strong for active mode to occur.  We selected the parameter values in Equation 8 by imposing the $T_{max}$-$g_{ij}$ relations for generating the quiescent- and active-mode time series that were presented earlier in this section.  Those relations were: $g_{ij}$($T_{max}$=0.5 mM) $=$ 0.01 nS, and $g_{ij}$($T_{max}$=1.8 mM) $=$ 2.0 nS, for quiescent and active mode, respectively.  The selected parameter values are: $T_{saturation}$ $=$ 2.0 mM, $\mu$ $=$ 0.18, and $\gamma$ $=$ 4.2.  

The time courses of $T_{max}$ and $g_{ij}$ that correspond to the path of Figure 6 are displayed in Figure 7.  The values of all parameters governing the $T_{max}$-$g_{ij}$ relation are listed in Table 6.
\setlength{\tabcolsep}{1pt}
\begin{table}[H]
\small
\centering
\begin{tabular}{ l c c|c c c c } \toprule
 \textit{Quantity} & Value  &   & \textit{Quantity} & Value &   \\\midrule 
 \textit{$T_{max}^{lowerbound}$} & 0.5 & mM & $\mu$ & 0.18 & -- \\
   &  &  &  &  &\\
 \textit{$T_{max}^{upperbound}$} & 5.0 & mM & $\gamma$ & 4.2 & -- \\
   &  &  &  &  &\\
 \textit{$T_{max}^{saturation}$} & 2.0 & mM & $\tau_r$ & 1. & ms \\
  &   &   & $\tau_f$ & 4. & ms \\\bottomrule
\end{tabular}\\
\caption{\textbf{Parameter values for the $T_{max}$-$g_{ij}$ relation.}  The lower bound on $T_{max}$ of 0.5 mM yields quiescent, but not active, FSU behavior.  The upper bound on $T_{max}$ was chosen to lie within the permitted range for generating active mode (see Figure 6).  $T_{saturation}$ was chosen so that the initial fast rise in $g_{ij}$ as a function of $T_{max}$, which was observed in simulations, would be obeyed.  Similarly, $\mu$ and $\gamma$ were chosen so that $g_{ij}$ assumes values of 0.01 and 2.0 nS for $T_{max}$ values of 0.5 and 1.8 mM, respectively.  The rise and decay constants $\tau_r$ and $\tau_f$ were chosen so that the time course mimics the time course during song: a rapid transition from quiescence to active state and a sustaining of the active state for a typical syllable duration of \textasciitilde 200 ms.} 
\end{table}

While our choice has some desirable features and allows us to probe the consequences of following a path in ($T_{max}$,$g_{ij}$) space, it is clear that the degrees of freedom here are numerous.  As noted, we plan to return to examine the possibilities.
 
\end{multicols}
\begin{figure}[H]
  \centering
  \includegraphics[width=190mm,valign=t]{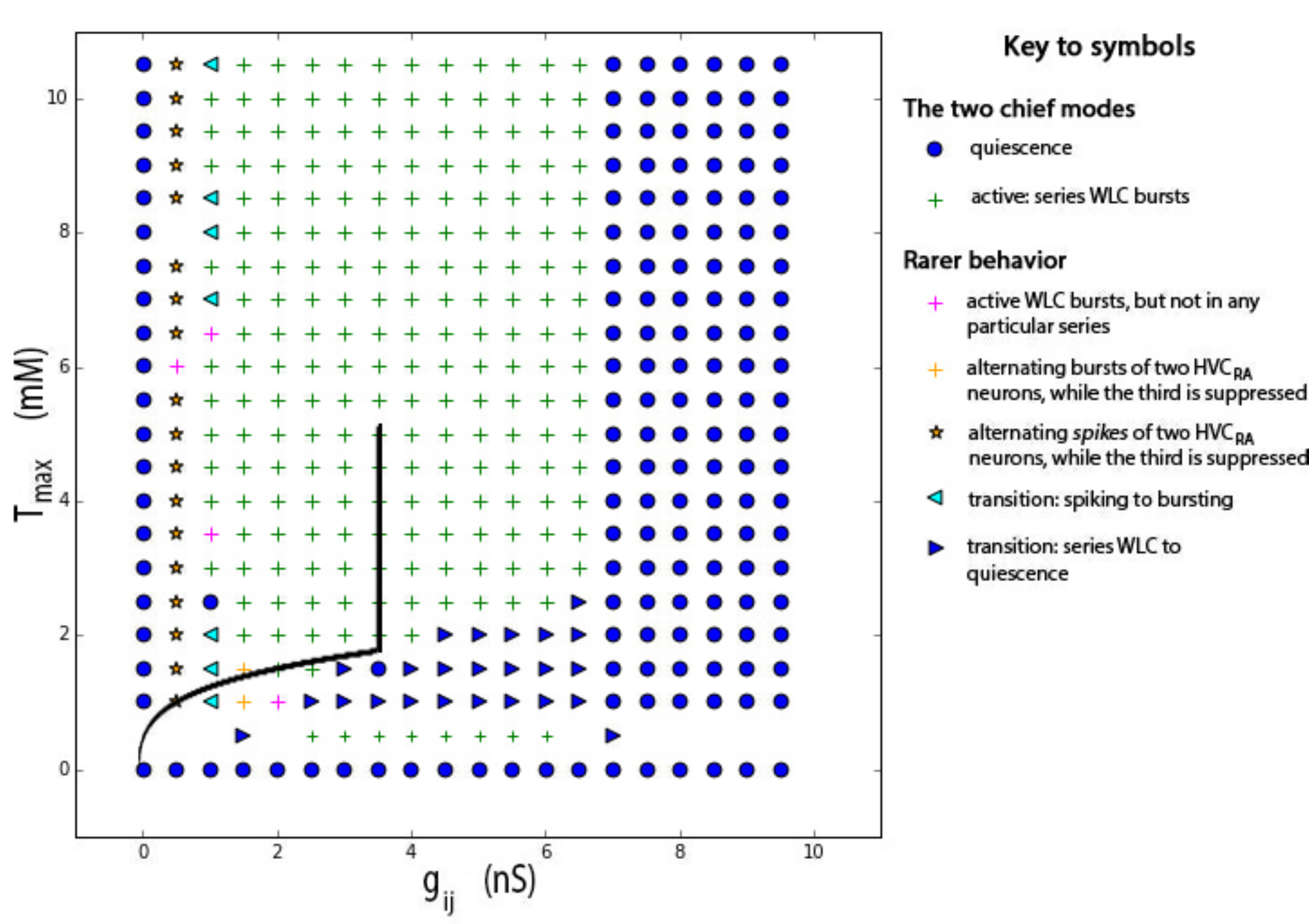}
  \caption{\textbf{FSU behavior as a function of $T_{max}$ and $g_{ij}$, where "behavior" is defined in terms of the $HVC_{RA}$ activity.}  The symbols defining behavior are listed at right.  The overlaid black path represents the trajectory for one neurotransmitter injection, as dictated by Equations 5-8.  The two chief modes (quiescence and active bursting series WLC) dominate the space and are robust to small changes in these parameter values.  In fact, the top row - at $T_{max}$ $=$ 10.5 mM - looks the same through a $T_{max}$ value of 50 mM.  In addition, there exist (at least) five modes of behavior within the transition regions between the two dominant modes, which one might expect to occasionally encounter in the laboratory.  The few locations on the grid that contain no symbol showed some combination of rarer modes and quiescence, and were difficult to characterize.}
\end{figure}
\begin{multicols}{2}
\begin{figure}[H]
\centering
  \includegraphics[width=95mm]{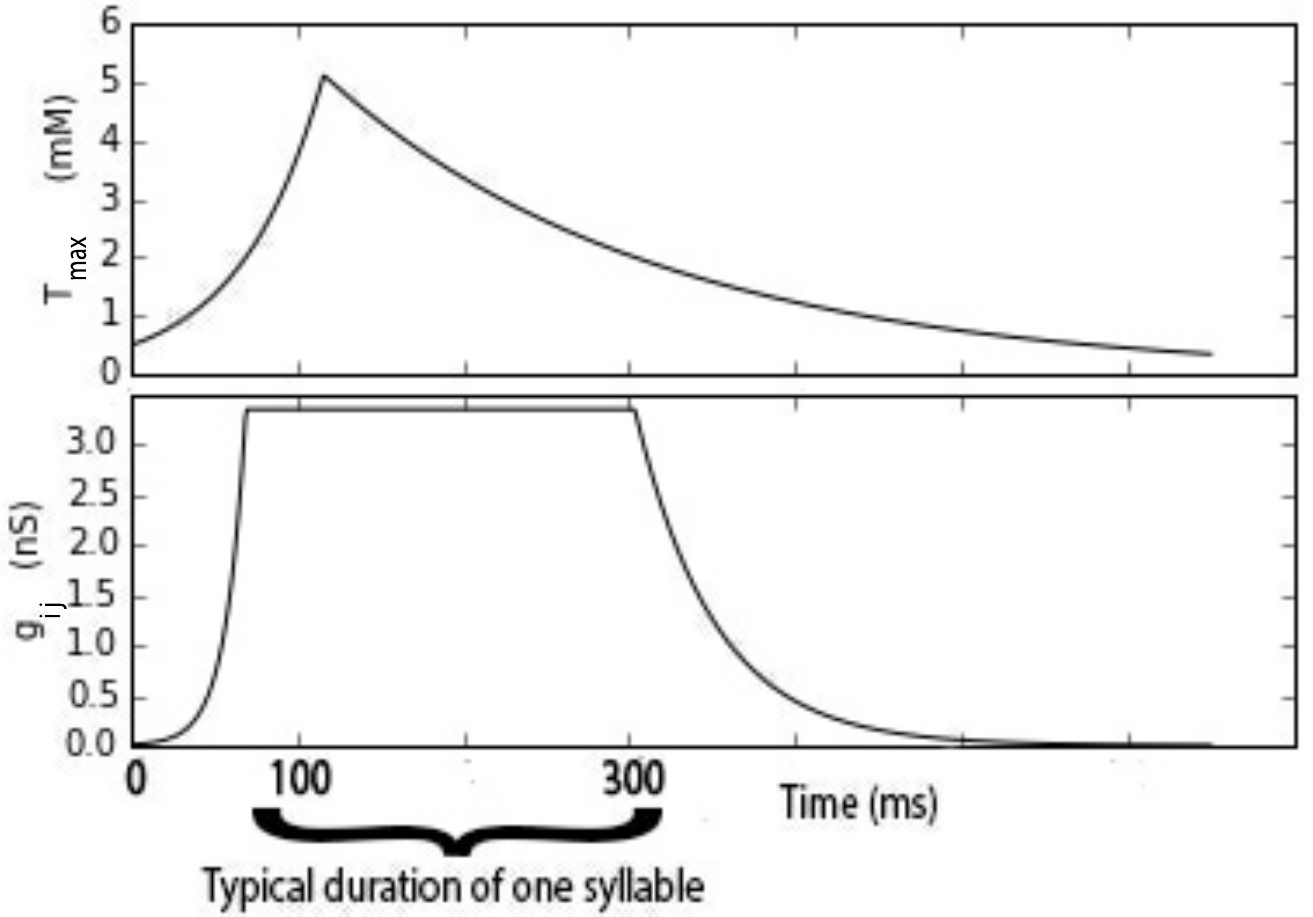}\\
  \caption{\textbf{The time course of neurotransmitter injection, in terms of  $T_{max}$(t) (\textit{top}) and $g_{ij}$(t) (\textit{bottom}) according to Equations 5-8}, and corresponding to the black path of Figure 6.} 
\end{figure}
We now return to our examination of the modes present in Figure 6.  As noted, in addition to the two dominant modes pictured in the diagram, we have identified five additional modes that inhabit a small but finite area.  Each of these "rarer modes" occurs reliably given the respective values of $T_{max}$ and $g_{ij}$ indicated in the diagram.  These modes are: i) alternating (winnerless competition) bursts, where the alternations occur in no particular repeating series (magenta plus signs); ii) alternating bursts of two out of the three $HVC_{RA}$ neurons, while the third is suppressed (yellow plus signs); iii) alternating \textit{single spikes} of two out of the three $HVC_{RA}$ neurons, while the third is suppressed (yellow stars); iv) transition from purely-spiking to purely-bursting activity (cyan left-pointing triangle); v) transition from series winnerless competition bursting to quiescence (blue right-pointing triangle).  

One might expect, then, to \textit{occasionally} encounter in the laboratory an $HVC_{RA}$ neuron exhibiting one of these alternative modes of behavior, rather than simply quiescence or simply a one-time burst.  In light of this discovery of additional modes, in the next section - \textit{Building Complete Songs} - we will examine an unusual and unexplained $HVC_{RA}$ voltage trace of Hahnloser et al. (2002).

\subsection{Importance of the excitatory-to-inhibitory connections}

In our simulations of an active FSU, we found an unexpected result regarding the excitatory-to-inhibitory connections (as pictured in Figure 1): the series of $HVC_I$ neuron firings cannot occur without them.  We discovered this requirement when, while maintaining the minimum value of 1.8 mM for $T_{max}$ and 2 $\mu$S for the inhibitory-to-inhibitory coupling strengths that are required for active mode, we lowered the excitatory-to-inhibitory connections from 1.0 to 0.01 nS.  The series of $HVC_I$ neuron firings then failed to occur.  Rather, one $HVC_I$ neuron continuously suppressed the other two.  We noted a similar failing of the series when we removed any one of the excitatory connections independently.  An increase in the inhibitory-to-inhibitory coupling strengths did not remedy this effect: similar results occurred for all values of inhibitory-upon-inhibitory connections from 1 to 20 $\mu$S.  We conclude: In our model circuit, mutual inhibition alone is insufficient to achieve winnerless competition; feedback from the excitatory cells is required.  We find this result to be extremely interesting, with the possible implication that variable concentrations of \textit{excitatory} neurotransmitters are also contributing to HVC dynamics (see \textit{Discussion}).  

\subsection{Model scalability}
  
Our FSU model is, in principle, generalizable to an arbitrary number of $HVC_I$ neurons and $HVC_{RA}$ ensembles.  For small networks (with neuronal populations on the order of tens to hundreds), the following rules work to effect the bimodal behavior that we have described: 
\begin{itemize}
  \item the all-to-all inhibitory-to-inhibitory connectivity must be preserved; 
  \item each $HVC_I$ neuron projects to all but one $HVC_{RA}$ ensemble; 
  \item each $HVC_I$ neuron "chooses" a distinct $HVC_{RA}$ ensemble to deny a projection; that is: no two $HVC_I$ neurons omit the same $HVC_{RA}$ ensemble.
\end{itemize}  
\noindent
To examine how the model might scale for networks of biological size is beyond the scope of this paper.  We also have yet to examine how the excitatory-to-inhibitory projections scale.\footnote{In addition, not all $HVC_{RA}$ neurons in one active FSU must receive a turn to fire.  We suspect that a given FSU contains a reservoir of $HVC_{RA}$ ensembles: significantly more than are required to generate one syllable in the trained bird.}
\end{multicols}  

\section{\\BUILDING COMPLETE SONGS}
\begin{multicols}{2}
We now demonstrate how multiple FSUs can be used to build a complete song.  We use as our example the highly stereotyped song "motif" of the zebra finch.  In doing so, we will show how the observed qualitative behavior of both excitatory and inhibitory populations in HVC during song are reproduced.

A trained male zebra finch sings a motif consisting of an invariant number of syllables separated by gaps corresponding to inhalation.  Each recorded $HVC_{RA}$ neuron is observed to fire once during each motif, and reliably at a particular temporal location.  Each $HVC_I$ neuron is observed to fire relatively continually throughout the song, with intermittent silences.   During normal, uninterrupted singing, the syllable order is invariant and both syllables and the silent inter-syllable gaps are precisely timed upon repeated renditions of the song.  This information is represented in the experimental raster plot by Hahnloser et al. (2002), which we have reproduced in Figure 8.    

Within the "FSU framework", the full motif \textit{could} be attributable to a chain-like propagation linking FSU to FSU within HVC.  In light of various lines of evidence that the syllables represent relatively independent structures in HVC, we seek another explanation.  Studies of learning in juveniles have shown that syllables become stabilized by inhibition independently as they are learned (Vallentin et al. 2016).  Attempts to interrupt song have indicated that individual syllables are relatively robust compared to the full motif: the motif can be interrupted by non-invasive techniques (Cynx 1990), but syllable interruption requires direct electrical interference (Ashmore et al. 2005; Vu et al. 2005).  Finally, KVL15 found that global GABA antagonist infusions to HVC degraded song while local GABA in-
\end{multicols}
\begin{figure}[H]
\centering
  \includegraphics[width=110mm]{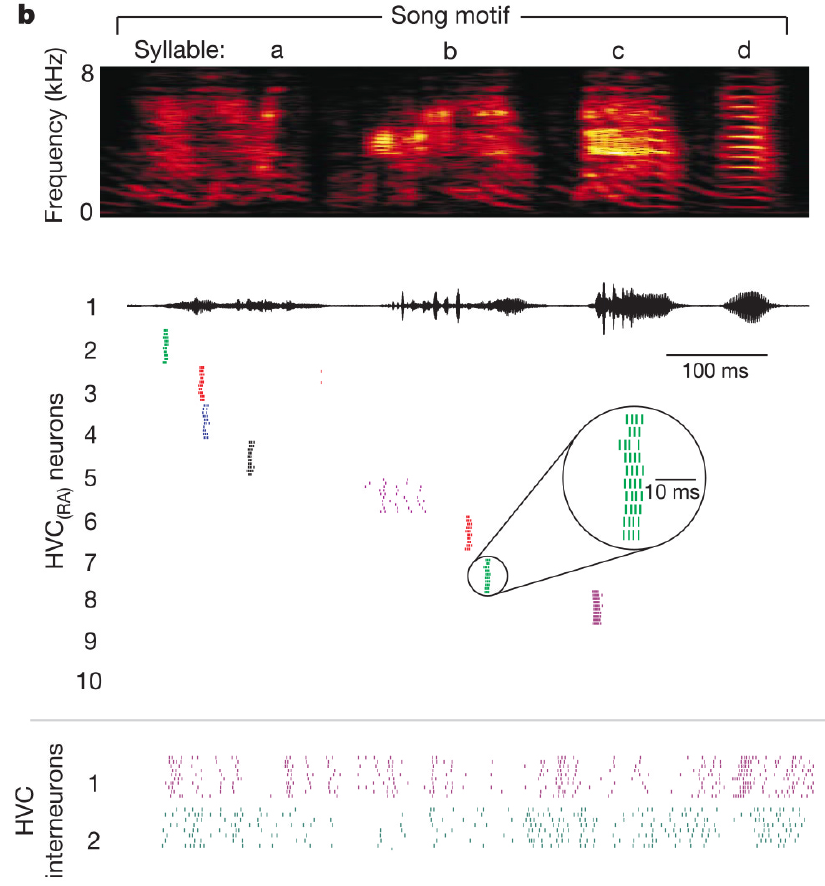}\\
  \caption{\textbf{A raster plot of spike times of $HVC_{RA}$ and $HVC_I$ neurons during repeated renditions of the zebrafinch motif, reproduced from Hahnloser et al. (2002)}.  The reader may find it of interest to compare this figure to the \textit{simulated} raster plot of Figure 9.  \textit{Reprinted by permission from Macmillan Publishers Ltd: Nature (Hahnloser et al. 2002).}} 
\end{figure}
\begin{multicols}{2}
\noindent
fusions affected the activity of certain cells in the region of infusion but the overall song was preserved.  These three lines of evidence suggest that, within our model framework, the FSUs function relatively independently.  In our example model of a zebra finch four-syllable motif, then, we take the inter-FSU connectivity to be essentially nonexistent and instead invoke a neural feedback loop as a mechanism for activations of a succession of four FSUs.

To simulate a full song consisting of four syllables, we sequentially exposed a set of four FSUs to an identical injection of neurotransmitter according to our formulation of the path described by Equations 5-8 (the black path of Figure 6), with the corresponding shapes of rise and fall of $T_{max}$ and $g_{ij}$ with time depicted in Figure 7.  The decay constant for $T_{max}$ ($\tau_f$ in Equation 7) was chosen so that the permitted duration of series WLC bursting activity is \textasciitilde 200 ms, a typical syllable duration.  This value permits time for all three $HVC_{RA}$ ensembles in one FSU to burst exactly once. 

Figure 9 shows the simulated raster plot that results when four unconnected FSUs are sequentially given the injection pictured in Figure 7.  At the very top, a sequence of four neurotransmitter injections is represented.  We assume that the first injection is initiated by an external stimulus, while the subsequent three are sustained by a neural feedback loop.  Just below the $T_{max}$(t) illustration at the top of Figure 9, the four corresponding FSUs are depicted, where purple electrodes labeled 1-10 have been attached by an experimenter to neurons whose identities are not known to the experimenter.  The first eight electrodes attach to $HVC_{RA}$ neurons; electrodes 9 and 10 attach to $HVC_I$ neurons.  

Below the FSU diagrams is the raster plot of firings of the ten neurons over the course of the song, where the vertical numberings (black) correspond to the numberings of the electrodes at top (purple).  Neuron 1, whose spike timings are green, corresponds to one $HVC_{RA}$ neuron in the "top" ensemble of FSU 1.  Neurons 2 and 3 correspond to different $HVC_{RA}$ neurons in the same "left" ensemble of FSU 1.  
\end{multicols}
\begin{figure}[H]
\centering
  \includegraphics[width=170mm]{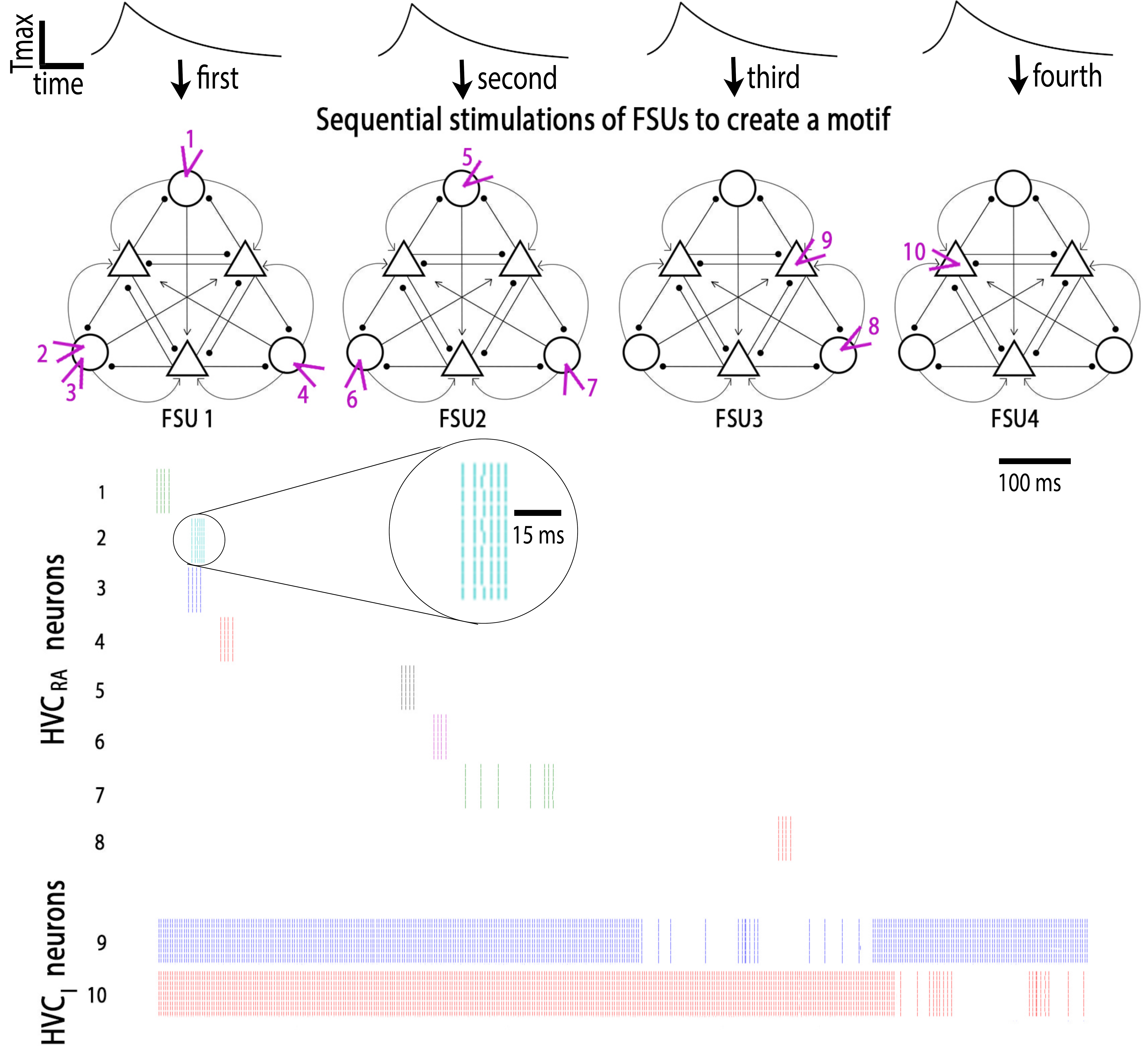}
  \caption{\textbf{A simulated raster plot of bursting $HVC_{RA}$ and spiking $HVC_I$ neurons during song.}  At top: four neurotransmitter injections sequentially target four FSUs.  Ten electrodes (purple arrowheads) have each been inserted into one neuron, by an experimenter who is blind to the neurons' identities.  The resulting action potential timings are shown below.  See text for important notes.}  
\end{figure}
\begin{multicols}{2}
We make three specific notes regarding the simulated raster plot.  First, the spike timings of Neurons 2 and 3 (cyan and blue, respectively) line up in nearly the same temporal window, but are offset slightly by different noise terms in the voltage time course of the somatic compartment for each neuron\footnote{To be clear: Each neuron was consistently given a different permitted noise range, which is why Neuron 3 consistently fires slightly prior to Neuron 2.}.  The reader might find it of interest to compare these simulated plots for Neurons 2 and 3 to the \textit{observed} raster plots of Neurons 2 and 3 in the plot of Hahnloser et al. (2002) (Figure 8).

Second, note the atypically long simulated spike train of Neuron 7, which belongs to FSU 2.  The rotation of activity around FSU 2 was timed to reach this neuron during the decay stage of $T_{max}$, such that FSU 2 entered one of the "rarer modes" that may occur within a transition region: the mode in which two $HVC_{RA}$ ensembles are spiking in alternation while the third is suppressed.  This mode is represented by a yellow star in Figure 6; note that the black path passes directly through it.  The reader may find it of interest to compare these simulated timings of Neuron 7 to the \textit{observed} timings of Neuron 5 in Hahnloser et al. (2002) (Figure 8).  Those authors found Neuron 5 to reliably exhibit activity that was not describable as a single burst, upon repeated song renditions.  To date, a satisfying explanation of that behavior has not been offered.  Our model predicts that activity reminiscent of that behavior - that is, neither strictly quiescence nor strictly active WLC - \textit{should} be observed in $HVC_{RA}$ neurons upon occasion.

Finally, note the behavior of the two $HVC_I$ neurons (9 and 10) at bottom.  For the majority of the song, when the FSU to which each $HVC_I$ belongs is  quiescent, the $HVC_I$ neuron spikes continually.  This is also the case for population activity of $HVC_I$ cells when the awake bird is not singing: the $HVC_I$ neurons are active continually and $HVC_{RA}$ neurons are essentially silent (Kozhevnikov \& Fee 2007).  Then, when the FSU to which a particular $HVC_I$ neuron belongs becomes active, there appear occasional lapses in the simulated activity of that $HVC_I$ neuron, consistent with observations.  

The songs of other species might be constructed via similar considerations of the relationships among syllables.  The starling sings a more complicated pattern that reflects a richer interplay among syllables.  A model of the starling song thus might embody more complicated connectivities among FSUs.  We look forward to investigating this possibility.  In general, for the song of any species, we expect that inter-FSU connectivity is significantly less extensive than within-FSU connectivity.
\end{multicols}

\section{\\DISCUSSION}
\begin{multicols}{2}
\subsection{Central-pattern-generator-like activity as a "winnerless competition"}
The observed phenomenon of patterned activity among inhibitory neurons has been explained in terms of an interplay between competition and background excitation.  The phenomenon was formalized in a framework called "winnerless competition" (WLC) by Rabinovich et al. (2001).  WLC has been used to model some biological circuits that display central-pattern-generator-like (CPG) behavior, including a pyloric circuit (Huerta et al. 2001), molluscan hunting behavior (Varona et al. 2002), and olfactory processes in locusts (Laurent et al. 2001).  

Rabinovich et al. (2001) describe a "node" as a cooperative ensemble of neural clusters in state space; they showed that one can alter the stabilities of nodes by altering the synapse strengths.  For low values of coupling strengths all nodes are stable attractors, corresponding to neurons (or ensembles) that may all be active simultaneously.  For higher coupling values, each node can correspond to a saddle fixed point, where a trajectory in state space is a closed heteroclinic orbit that sequentially traverses stable limit cycles in the vicinity of each saddle fixed point.  This configuration represents sequential switching of activity among the three nodes.  (See Rabinovich et al. 2013 for pictorial illustrations of these two modes of activity.)  The neuronal interactions are defined in terms of competitive Lotka Volterra dynamics.  (See Zeeman 1993 for a bifurcation analysis of a competitive three-dimensional Lotka Volterra system, and Afraimovich et al. 2004 for a phase space analysis of a "winnerless" Lotka Volterra system).  

The dynamics of our model are consistent with this framework, save one feature: the reliance of stable series activations on feedback from the excitatory $HVC_{RA}$ neurons.  In our simulations, this feedback was a necessary condition for series propagation in an active FSU, in addition to threshold values of $T_{max}$ and inhibitory-to-inhibitory couplings.  Such dynamics are not considered within the context of WLC theory.  We are uncertain of the biophysical implications of this feature; we suspect that the excitatory feedback serves to stabilize the competitive mode of behavior so that patterns of alternating activity can occur.   

\subsection{Plausibility of a chemically-delivered signal to alter synaptic coupling strengths}
  
We have suggested that a bird's need to vocalize leads to an injection of inhibitory neurotransmitter in the vicinity of an FSU, which in turn initiates the first syllable of song.  We have further suggested that a feedback loop triggers sequential releases of neurotransmitter, to play a complete song.  Here we examine the plausibility of such a neuromodulatory mechanism and its timescale of action.  \\

\noindent
\textit{Sources of neuromodulation}\\

Sources of transient changes in synaptic action (commonly called "synaptic agonist transients" or "transients") have been identified throughout the central nervous systems of many species.  In some mammalian brains, the ventral tegmental area (VTA) has been identified as a reservoir of dopaminergic cells (Phillipson 1979) that, in the avian brain, project to and impose neuromodulatory effects upon regions throughout the CNS and are a common source of transients.  The VTA mediates a variety of tasks in the brain, including inhibitory action in some locations (e.g. Stamatakis et al. 2013).  While direct projections from VTA to HVC in the avian brain have not been identified, there does exist some evidence for a role of VTA in mediating birdsong: multi-neuron activity in the VTA of Bengalese finches has been found to consistently increase prior to the initiation and termination of song bouts (Kapur 2008).  We consider it reasonable to postulate that VTA affects HVC, even indirectly, as the VTA serves numerous brain areas in a neuromodulatory role.  We do not make any claim that VTA is involved in HVC modulation, but rather we suggest that it merits an examination. \\

\noindent
\textit{Mechanisms of neuromodulation and effects on synapses}\\

Both the mechanisms of neuromodulatory action and their effects on synapses are difficult to study experimentally, due to the inaccessibility of narrow synaptic clefts, uncertainties in the geometry of the synapse structure, and the observation that synapses vary widely in geometry (Scimemi \& Beato 2009).  Rise times are coupled to receptor dynamics; decay times can be mediated by diffusion, reuptake, binding to receptors, and enzymic breakdown.  

The magnitude over which a neurotransmitter concentration can change in a synaptic cleft varies widely.  Most sources cite a saturating value on the order of 1 mM before dropping over two to three orders of magnitude (Scimemi and Beato 2009; Barberis et al. 2011).  The effects of such changes on synaptic coupling is generally measured via current responses (e.g. Mozrzymas 2004), but no direct methods are available to measure the transient at synapses.  

For a coarse idea, we look to behavioral studies.  Here, we implicitly assume that a change in macroscopic behavior reflects a change, on the cellular level, in circuit modality.  Much work in this area has focused on dopamine as a modulator of behavior.  Dopamine levels in Area X of zebra finches have been found to be significantly lower during undirected compared to sexually-motivated song (Heimovics \& Riters 2008), which suggests that dopaminergic neurotransmission may differentially modulate vocal behavior depending on context.  

We distill from this information that it is reasonable to propose neuromodulation as a mechanism for transitioning between a circuit's modes of activity, and that the mechanism may involve a change of synaptic coupling strengths.  We further infer that our model's requirement of a roughly threefold increase in $T_{max}$ to effect such a change falls well within the bounds set by the current level of understanding. \\

\noindent
\textit{Timescales of neuromodulatory action}\\

Burst times of $HVC_{RA}$ neurons during song have a typical precision of \textasciitilde 1 ms (Kozhevnikov \& Fee 2007).  Our model thus implies that neurotransmitter rise and decay timescales must be that precise if neuromodulation is to serve as the mode-switching mechanism.  We also have implied that the temporary elevation of neurotransmitter can sustain an FSU's active mode for hundreds of milliseconds, a typical syllable duration.  How reasonable are these requirements?

Rise times of neurotransmitter concentrations on the order of milliseconds have been reported (e.g. Robinson et al. 2003).  In addition, routine injected neurotransmitter "pulses" that are designed to simulate real synaptic action typically are delivered over 0.1 ms - implying that neurotransmitter concentrations in real biological circuits are believed to be capable of changing significantly over 0.1 ms.  

How long can a transient's effect on a circuit last?  The time course can vary from one to hundreds of milliseconds, depending on the magnitude of the initial injection and other modulatory factors such as the presence of particular receptors.  Scimemi \& Beato (2009), for example, showed that the duration of post-synaptic glutamatergic currents can vary over two orders of magnitude, depending on whether they are mediated by AMPA receptors alone ($t_{decay}$ \textasciitilde 2 ms) or if NMDA receptors are also recruited ($t_{decay}$ \textasciitilde 200 ms).  

Finally, we are concerned about the temporal precision of neurotransmitter decay.  Can such a process occur reliably with a precision of 1 ms, upon the sounding of each syllable?  The consideration here is not the shape of neurotransmitter fall-off, but rather that the fall-off reach one critical value at a reliable time following injection.  That is: in our model, the series that occurs during active mode does not slow gradually as neurotransmitter concentration decreases.  Rather, the series occurs at a constant rate above the threshold concentration, below which there occurs a sharp transition back to quiescence.   An answer to our question here would be speculative, however, given the current experimental uncertainties.

\subsection{Possible use of excitatory-to-excitatory connections}

Some monosynaptic connections between $HVC_{RA}$ neurons have been identified experimentally (KVL15), although at a rate roughly 100 times lower than that of the reciprocal inhibitory-excitatory connections.  While our model does not require excitatory-excitatory monosynaptic connections, it does allow for them.  Here we note how excitatory-excitatory connections might arise within our FSU framework, and we offer a possible biophysical use for such connections.

We have identified two possible locations for excitatory-excitatory connections in our model.  One is within a particular ensemble of an FSU.  Such connections do not strike us as important, however, as unconnected $HVC_{RA}$ neurons in one ensemble may be active simultaneously even if not directly connected.  A second - and possibly useful - location for such connections could be: \textit{between} FSUs. 

Consider two FSUs, where there exists a unidirectional monosynaptic connection from FSU 1 to FSU 2.  Imagine that FSU 1 lies within the spatial region receiving a neurotransmitter injection, while FSU 2 does not.  When FSU 1 receives the injection, its ensembles may now participate in the corresponding syllable.  Given the connection to FSU 2, the ensembles of FSU 2 may also be recruited to participate in the syllable - even though FSU 2 has not received the injection directly.  In this way, excitatory-to-excitatory connections \textit{might} serve to reduce the required spatial extent of the injection.  This suggestion constitutes little more than speculation; we offer it in order to address the observed low rate of such connections.
 
\subsection{Attempts to incite FSU mode-switching via electrical manipulation}

We describe briefly an alternative method we investigated to achieve an active series of firings of $HVC_I$ neurons: electrical stimulation.  Electrical signaling can occur faster than chemical signaling, and on a timescale whose precision has been established well beyond the current state of understanding of the temporal precision of neuromodulatory action.  We initially considered it to be the most likely candidate for an FSU mode-switching mechanism.

Beginning from the quiescence regime, we attempted to incite series activity via current pulses delivered to particular cells in the circuit.  We gave a current pulse (shorter than 10 ms) atop the background current to various subsets of neurons in one FSU.  The pulse-receiving neuron(s) spiked once in response; the other neurons did not respond to the pulse for any biophysically realistic values of injected current (lower than 700 pA) or synapse strength (lower than 100 $\mu$S).  The circuit activity resumed its previous behavior within 10 ms.  We concluded that if electrical stimulation is to effect mode-switching, more elaborate injection designs would be required.  Seeking a simpler solution, we looked elsewhere.

\subsection{The silent inter-syllable gaps}

We chose not to include the silent inter-syllable gaps in our model of HVC, because at this time we find no particularly compelling representation for a gap.  We suspect, however, that the gaps are integral to song production and have a temporal representation at the level of HVC.  Cooling HVC uniformly stretches syllables and gaps while cooling the nucleus RA had no effect on song (Long \& Fee 2008).  More importantly, correlations between respiratory action and activity in HVC have been demonstrated (Andalman, Foerster, \& Fee (2011); Amador et al. (2013)).  We speculate that the duration of each silence conveys part of the informational content of song.

\subsection{Structural versus functional connectivity rates}

We comment briefly on relating our model's functional connectivity to structural connectivity rates, where functional connections are defined as the subset of structural connections that are currently in use by the circuit.  Our model requires high rates of functional connectivity within an FSU.  We would like to directly compare these rates to observed structural connectivity rates (e.g. Mooney \& Prather (2005); KVL15).  Such a comparison, however, would require knowledge of the learning stage (see Okubo et al. (2015) for evidence of synapse pruning during learning in HVC) and the rate at which a formerly-used synaptic connection dissolves (see Luo \& O'Leary (2005) and Walsh \& Lichtman (2003) for evidence that axons projecting to unused synapses have some timescale of retraction or degeneration).  Calcium imaging techniques appear to hold the greatest promise for illuminating the relationship between structural and functional connections.  Until then, we will withhold speculation. 

\subsection{Spatial considerations}

The spatial organization of FSUs, which we have not considered in this paper, is an important topic to consider.  There exists evidence for such organization in HVC (Stauffer et al. 2012, Day et al. 2013, Poole et al. 2012), and efforts are underway to incorporate evidence of spatial organization into a model of functional connectivity (Markowitz et al. 2015).

\subsection{A concern regarding model robustness to small differences among synapse strengths}

As noted, in our numerical simulations the relative values of the six interneuron-to-interneuron $g_{ij}$ values must be identical to one part in 20 in order for the series WLC activity to occur reliably.  This is a worrisome weakness.  In a follow-up paper we will examine possible reasons for this sensitivity and possible methods to remedy it.

\subsection{Model predictions and suggested experiments}

Here we suggest experimental tests of i) the FSU model predictions, and ii) the suggestion that neuromodulatory action initiates song.  These tests can currently be performed in the laboratory.  \\

\noindent
\textit{Testing FSU model predictions}\\

\noindent
The plot of FSU behavior as a function of $T_{max}$ and $g_{ij}$ (Figure 6) reveals five additional modes of behavior that reside within the slender transition regions between quiescent and active mode.  This finding predicts that one should occasionally find $HVC_{RA}$ neurons exhibiting activity that is neither strictly below action potential threshold nor strictly a one-time-burst event.  Furthermore, such atypical behavior of a particular neuron may be expected to occur reliably at a particular temporal location during song.  Hahnloser et al. (2002) \textit{might} have encountered such an event, depicted by their raster plot of Neuron 5 (Figure 8).  A deeper examination of this possibility requires the recording of a significantly larger number of $HVC_{RA}$ neurons during song.  \\

\noindent
\textit{Testing the proposal that neuromodulation initiates song}\\

To examine whether neuromodulation plays a role in initiating song, it might be helpful to examine whole-cell recordings in the awake bird, where - in place of an external stimulus - a neuromodulatory agent is delivered to HVC.  The aim here is to identify some neuromodulatory agent that can act as the external stimulus, to initiate song.  Heimovics \& Riters (2008) performed such an experiment in the starling HVC, and found that infusing dopamine agonists stimulated song while antagonists hindered it.  This experiment merits repeating.  Furthermore, it would be instructive to compare the result of infusing such a neurotransmitter both globally and locally to HVC.  In our model, a global infusion would ignite all FSUs in HVC simultaneously; whether it would be possible for patterned activity to arise in that scenario, we would be interested to learn.  

A second suggestion is to specifically target VTA, with experiments similar to the work of Kapur (2008).  That author found that VTA activity is associated with the onset and offset of song bouts.  The nature of that association was not determined, and the finding merits a follow-up examination.  As noted earlier, we are making no claim that VTA is the probable origin of the process that modulates $T_{max}$ in HVC.  Rather, as VTA has been identified as a modulator of numerous brain areas, we are suggesting that it would be worthwhile to examine its possible role in the scenario presented in this paper.

\end{multicols}

\section{\\A LOOK FORWARD}
\begin{multicols}{2}

We have offered a biophysically-based model of a fundamental bi-modal unit of song representation in the avian HVC.  The two modes are robust to small variations in values of the parameters governing their dynamics.  When considering a population of such FSUs, much of the experimentally observed population activity of $HVC_I$ and $HVC_{RA}$ neurons can be reproduced qualitatively.  Additional modes of activity can occur during the transition regions between the two dominant modes, and on occasion these should be identifiable in the laboratory.  We have also discovered that excitatory feedback to the inhibitory population is necessary for a stable propagation of series firings within the framework of winnerless competition - a finding that calls for a physical explanation.  

We have offered no details of how the FSU represents the temporal and spectral content of a syllable.  Could the rate of series propagation be related?  Could the number of $HVC_{RA}$ ensembles firing per syllable be related?  Both possibilities might be addressed by performing tests akin to the raster plot of Hahnloser et al. (2002) during an experiment in which the ambient temperature is varied.  Such "cooling studies" have revealed direct relationships among temperature, rate of song production, and spectral content (e.g. Long \& Fee 2008).   

Many questions arise regarding our proposed mechanism for effecting a bi-modal FSU.  If an injection of inhibitory neurotransmitter is responsible, then: By what mechanism is the "donor" inhibitory cell triggered to release neurotransmitter?  Is this donor cell internal or external to HVC?  How does the neurotransmitter injection find its target FSU?  What spatial precision is required of the injection?  It would be fascinating to ultimately engineer an experiment capable of targeting individual FSUs with a neurotransmitter injection.  It would serve as an ultimate test of our model if, for example, in the zebra finch brain the "FSU 2" of Figure 9 could be reliably identified - and reliably ellicit the bird's "Syllable B" repeatedly, upon repeated targeted injections to that FSU. 

Finally, there is the critical question of model robustness on large scales.  How strictly must our "scalability rules" be followed by a circuit consisting of thousands of cells and synaptic connections?  Do the rules permit some reasonable degree of flexibility?  In follow-up papers we will address these questions and describe a method of statistical data assimilation that can, in principle, determine whether a proposed functional connectivity accurately represents data from a real biophysical circuit.
\end{multicols}  

\section{\\ACKNOWLEDGEMENTS}
\begin{multicols}{2}
We appreciate Arij Daou, Daniel Margoliash, and Michael Long for enlightening explanations, perspectives, and translations.  Misha Rabinovich, Daniel Breen, and Nirag Kadakia offered useful conversations.  Thanks also to three anonymous reviewers for valuable feedback on an earlier version of this paper.  This research was funded by the Office of Naval Research grant N00014-13-1-0205.  
\end{multicols}
\bibliographystyle{acm}
\nocite{*}
\bibliography{bibliography2}

\end{document}